\documentclass[journal]{IEEEtran}
\usepackage{amsmath,graphicx}
\usepackage{amsfonts}
\usepackage{enumerate}
\usepackage{algorithm} 
\usepackage{algorithmic} 

\newtheorem{theorem}{Theorem}

\usepackage{caption}

\usepackage{mathrsfs}
\usepackage{stfloats}
\usepackage{enumerate}
\usepackage{epstopdf}
\usepackage{subfigure}
\usepackage{float}
\usepackage{amsbsy}
\usepackage{color}

\normalsize

\ifCLASSINFOpdf

\else

\fi

\begin{document}

\title{Double Intelligent Reflecting Surface-assisted Multi-User MIMO mmWave Systems with Hybrid Precoding}

\author{Hehao Niu, Zheng Chu, \emph{Member, IEEE}, Fuhui Zhou, \emph{Senior Member, IEEE}, Cunhua Pan, \emph{Member, IEEE}, \\Derrick Wing Kwan Ng, \emph{Fellow, IEEE}, and Huan X. Nguyen, \emph{Senior Member, IEEE}
        \thanks{

        }

}

\markboth{IEEE xx,~Vol.~xx, No.~xx, ~2021}%
{Niu \MakeLowercase{\textit{et al.}}: Double Intelligent Reflecting Surface-assisted Multi-User MIMO mmWave Systems with Hybrid Precoding}
\maketitle

\begin{abstract}
This work investigates the effect of double intelligent reflecting surface (IRS) in improving the spectrum
efficient of multi-user multiple-input multiple-output (MIMO) network operating in the millimeter wave (mmWave) band. Specifically, we aim to solve a weighted sum rate maximization problem by jointly optimizing the digital precoding at the transmitter and the analog phase shifters at the IRS, subject to the minimum achievable rate constraint. To facilitate the design of an efficient solution, we first reformulate the original problem into a tractable one by exploiting the majorization-minimization (MM) method. Then, a block coordinate descent (BCD) method is proposed to obtain a suboptimal solution, where the precoding matrices and the phase shifters are alternately optimized. Specifically, the digital precoding matrix design problem is solved by the quadratically constrained quadratic programming (QCQP), while the analog phase shift optimization is solved by the Riemannian manifold optimization (RMO). The convergence and computational complexity are analyzed. Finally, simulation results are provided to verify the performance of the proposed design, as well as the effectiveness of double-IRS in improving the spectral efficiency.
\end{abstract}

\begin{IEEEkeywords}
Intelligent reflecting surface, mmWave communications, hybrid precoding, majorization-minimization, Riemannian manifold optimization.
\end{IEEEkeywords}

%

\IEEEpeerreviewmaketitle
\setlength{\baselineskip}{1\baselineskip}
\newtheorem{fact}{Fact}
\newtheorem{assumption}{Assumption}
\newtheorem{lemma}{\underline{Lemma}}[section]
\newtheorem{corollary}{\underline{Corollary}}[section]
\newtheorem{proposition}{\underline{Proposition}}[section]
\newtheorem{example}{\underline{Example}}[section]
\newtheorem{remark}{\underline{Remark}}[section]
\newcommand{\mv}[1]{\mbox{\boldmath{$ #1 $}}}

\section{Introduction}
For the past decades, multiple-antenna techniques have attracted great interest, since they can increase the spectrum efficient \cite{Wong2017}. However, the fully digital beamforming (BF) or the precoding structure imposes an excessive energy consumption and hardware cost due to the use of various number of radio frequency (RF) chains as well as the analog-digital (A/D) converters \cite{ZhangJSAC20201}. To handle this issue, the hybrid A/D BF or hybrid A/D precoding technique has been regarded as an efficient approach which requires only a smaller number of RF chains than the fully digital counterpart \cite{IoushuaTSP2019}. For instance, in \cite{AroraTSP2020} and \cite{GongTSP2020}, the majorization-minimization (MM) based methods were proposed to design the hybrid BF/precoding in multiple-input single-output (MISO) and multiple-input multiple-output (MIMO) networks, respectively. Also in \cite{XingTSP2019}, a matrix monotonic optimization-based hybrid precoding design was developed. Furthermore, an alternating optimization (AO) approach by exploiting the orthogonal property of the digital BF was presented in \cite{LinTC2019}. However, in spite of the fruitful research in related literatures, the performance of millimeter-wave (mmWave) communication systems is still sensitive to blockages which hinders the communication reliability \cite{AyachTWC2014}--\cite{3GPP}. Hence, there is an emerging need for a new technology to solve these problem.

Recently, a new technique called intelligent reflecting surface (IRS), has inspired great research attention. To be specific, an IRS comprises an array of reflecting elements, which can reflect and alter the phase of the electromagnetic (EM) wave passively. Hence, by smartly tuning the phase shifts with a programmable controller, the reflected signals can be adjusted to a desired direction \cite{TangJSAC2020}. Moreover, since only reflecting the received signal without a dedicated RF processing, or re-transmission, IRS can have higher energy efficiency than active transmitter (Tx) or relay \cite{WuTC2021}. With these advantages, IRS has been treated as a promising way to overcome above-mentioned issues in mmWave systems. Currently, IRS has been applied to various practical settings such as the MISO network in \cite{ZhouTSP2020,DuTWC2021}, the MIMO network in \cite{ZhangJSAC2020}-\cite{PanJSAC2020}, the cognitive radio (CR) network in \cite{ZhangTVT2020}, the physical layer security communication in \cite{NiuTC2021,HuTC2021}, the green communication network in \cite{YuTC2021}, and the two-way communication network in \cite{WangTC2021}, respectively. Among relevant works, MM and manifold optimization (MO) are two promising methods which have been widely used to address the formulated problem. Interested readers can refer to Appendix C for more details about MM and MO methods.

The above works commonly assumed that perfect channel state information (CSI) can be obtained. In fact, it maybe hard to obtain the perfect CSI, especially for the IRS-related links. Thus, the robust design has been investigated in IRS-aided system. For instance, \cite{ZhouTSP20202} studied the robust transmission design for an IRS-assisted MISO network, where a constrained concave-convex procedure (CCCP) algorithm was proposed. Also, in \cite{WangTWC20211}, the authors proposed an energy efficient BF design for IRS-empowered MISO networks, where a semidefinite relaxation (SDR)-based robust method was presented. Then, in \cite{YuJSAC2020}, the authors studied the robust and secure communications in IRS-aided MIMO network, where a penalty-based SDR method was developed. Moreover, \cite{YuanTC2021} studied the robust design in IRS-assisted CR network, where an AO-based method was proposed.

Furthermore, to exploit the potential of IRS in mmWave communication systems, in \cite{PradhanWCL2020}, the authors studied the hybrid precoding for IRS-enabled mmWave system, where a gradient-projection approach was presented. Then, in \cite{XiuTWC2021}, the authors investigated an IRS-enhanced mmWave non-orthogonal multiple access (NOMA) network, where a MO based method was developed to design the power allocation, phase shifters, and the hybrid BF. Also, in \cite{DiJSAC2020}, the hybrid BF and phase shift design in mmWave multi-user MISO channel was studied. In particular, IRS can be seen as an effective method to achieve hybrid BF/precoding since the analog phase shift can conduct by the reflection elements at the IRS. Recently, \cite{ZhaoTWC2021} studied the joint BF and user association optimization for IRS-assisted mmWave multiuser MISO networks. Besides, \cite{WangTWC2021} and \cite{Hong2021} investigated the joint precoding and phase shifts design for IRS-assisted mmWave massive MIMO systems. In fact, as pointed out in \cite{Lu2020}, IRS-enabled hybrid precoding/BF architecture is more energy-efficient than the conventional phased array based hybrid precoding/BF architecture, owning to the use of the energy-efficient IRS rather than the energy-hungry phased array at Tx.

The above works mainly studied the single-IRS scenario, to further enlarge the coverage for wireless communication systems, especially in high frequency such as mmWave band, on the other hand, some recent works have studied double-IRS-aided communications. To be specific, the cascaded passive BF design in a double-IRS-assisted network was investigated in \cite{HanWCL2020}. While a routing technique in a multi-hop IRS-enabled network was proposed in \cite{MeiWCL2021}. Then in \cite{ZhengTWC2021}, the authors proposed a phase shifter method in a double-IRS-enhanced uplink MIMO network. Also in \cite{Han2021}, for a single-user MIMO downlink network, under the assumption that LoS channel model for the inner-IRS link, the authors proved that double-IRS can obtain two times of the capacity scaling order than that of the single-IRS counterpart, thanks to the cooperative BF gain originated by the double-reflection link. Then, \cite{DongTVT2021} studied the secure transmission in double-IRS-empowered wiretap channel, where a product manifold algorithm was developed to optimize the secrecy rate. Moreover, in \cite{Zhang2020}, a MM based approach was proposed to handle the weighted sum rate (WSR) maximization design in a multi-hop IRS-aided network. The results in these works suggest the enormous potential of double-IRS in improving the spectrum efficiency and expanding the coverage of wireless network.

Motivated by these observation, in this work, we propose a new hybrid precoding structure in a multi-user downlink MIMO network aided by two cascaded IRSs, where the digital precoding is achieved by Tx and the analog phase shift is conducted by the IRS. To solve the formulated WSR design subject to the quality-of-service (QoS) constraints, we turn the objective function into a quadratic form by utilizing the MM method, then, a suboptimal block coordinate descent (BCD) method is developed to obtain the solution. Specifically, we propose a quadratically constrained quadratic program (QCQP)-based method to design the digital precoding matrix. Then, the analog phase shift is solved by a price-based MO algorithm. Simulation results are provided to validate the performance of the proposed approach. Our main contributions are concluded as follows:
\begin{itemize}
\item[1)]
Due to the sophisticated objective function and coupled variables, the WSR maximization problem is non-convex. To handle this issue, we develop an MM-based method to transform the objective function into an equivalent tractable quadratic form. Then, a BCD framework is applied to handle the reformulated problem, where each subproblem is solved by the corresponding method in an iterative manner. Comparing with the weighted minimum mean squared error (MMSE) method, the proposed MM-based method does not require prior knowledge about the number of the information streams for each user \cite{NaghshTSP2019}. Besides, the MM method can tackle the QoS constraints, e.g., the minimum rate constraints, which is more suitable for the formulated problem than the weighted MMSE method.
\item[2)]
Given the phase shift matrices, the digital precoding matrix at the Tx is optimized by the QCQP. As for the optimization of the phase shifters, we firstly propose an approximation of the non-smooth QoS constraints by using the log-sum-exp inequality, which is smooth and differentiable. Then, by including the approximated QoS constrains into the objective, we transform the phase shifters optimization problem into a quadratic form objective with the unit modulus constraint (UMC). Moreover, we develop a price mechanism-based Riemannian manifold optimization (RMO) algorithm to design the phase shift matrices, where the price factor can be found the by the bisection search method.
\item[3)]
The convergence of the proposed algorithm is guaranteed by rigorous proof. Besides, the proposed algorithm enjoys the polynomial time complexity, which is beneficial to implementation. Finally, simulation results are provided to validate the performance of the proposed method, as well as the effectiveness of double-IRS in improving the spectral efficiency in MIMO network.
\end{itemize}

The rest of this work is organized as follows. Section \ref{secSM} describes the system model and formulates the problem. Section \ref{method} investigates the joint design. Section \ref{secSimulations} provides the simulation results and Section \ref{secConclusions} concludes the work.

\textit{Notations:}
Throughout this work, the upper case boldface letters and lower case boldface letters denote matrices and vectors, respectively. \({\rm{Tr}}\left( {\bf{X}} \right)\), $\ln \left| {\bf{X}} \right|$, and ${\rm{rank}}\left( {\bf{X}} \right)$ represent the trace, the logarithmic determinant, and the rank of \(\bf{X}\), respectively. Besides, \({{\bf{X}}^T}\), \({{\bf{X}}^*}\), \({{\bf{X}}^H}\), ${{\bf{X}}^{ - 1}}$, and \({\rm{vec}}\left( {\bf{X}} \right)\) denote the transpose, the conjugate, the conjugate transpose, the inverse, and the vectorization of \(\bf{X}\), respectively. The block-diagonal matrix with diagonal entries ${{\bf{X}}_1}, \ldots ,{{\bf{X}}_N}$ is denoted by ${\rm{BLKDiag}}\left( {{{\bf{X}}_1}, \ldots ,{{\bf{X}}_N}} \right)$, and is reduced to ${\rm{Diag}}\left( {{x_1}, \ldots ,{x_N}} \right)$ when scalar diagonal entries are considered. ${\left[ {\bf{X}} \right]_{p,q}}$ denotes the entry at the $p$-th row and $q$-th column of ${\bf{X}}$, and ${\left[ {\bf{x}} \right]_m}$ denotes the $m$-th entry of ${\bf{x}}$, respectively. ${\rm{diag}}\left( {\bf{X}} \right)$ means a vector consists of the entries on the main diagonal of ${\bf{X}}$. In addition, ${\bf{I}}$ stands for an identity matrix and \({\bf{X}}\succeq{\bf{0}}\) indicates that \({\bf{X}}\) is positive semi-definite. $\left|  \cdot  \right|$, \(\left\|  \cdot  \right\|\), and \(\Re \left\{  \cdot  \right\}\) denote the modulus, the Frobenius norm, and the real part of a variable, respectively. Besides, $\mathbb{E}$ denotes the expectation, and \({\bf{x}} \sim {\cal C}{\cal N}\left( {{\boldsymbol{\sigma}},{\bf{\Sigma}} } \right)\) suggests that \({\bf{x}}\) is a circularly symmetric complex Gaussian random variable with mean \({\boldsymbol{\sigma}}\) and covariance \({\bf{\Sigma}}\).

\section{System Model and Problem Formulation}\label{secSM}
\subsection{IRS Model}
Firstly, we provide the IRS model. The $m$-th reflecting coefficient (RC) is denoted as ${\theta _m} = {e^{j{\phi _m}}}$, with ${\phi _m} \in \left[ {0,2\pi } \right)$. Thus, for an IRS with $M$ reflection elements, the RC matrix is given as ${\bf{\Theta }} = {\mathrm{Diag}}\left( {{\theta _1}, \ldots ,{\theta _M}} \right)$. Commonly, there exist two models to describe the RC.

{\bf{1) Continuous RC:}} In this case, the amplitude of the incident signal is unchanged, e.g., ${\left| {{\theta _m}} \right|} = 1$. While the phase can take any value, e.g.,\vspace{1.25ex}
\begin{equation}\label{eq:CPS}
{\theta _m} \in {{\cal F}_1} \buildrel \Delta \over=\left\{\theta_{m} | \theta_{m}=e^{j \phi_{m}}, \phi_{m} \in[0,2 \pi)\right\}.\vspace{1.25ex}
\end{equation}

{\bf{2) Discrete RC:}} In this case, the RC of each reflection element can only take a value from a discrete set. Let ${Q_\phi}$ denote the number of bit resolutions for each IRS element. Here, we assume that the phase shift values are equally taken in the region $\left[ {0,2\pi } \right)$, e.g.,\vspace{1.25ex}
\begin{equation}\label{eq:amppha}
{\theta _m} \in {{\cal F}_2} \buildrel \Delta \over = \left\{ {{\theta _m}\left| {{\theta _m} = {e^{j{\phi _m}}},{\phi _m} \in {\cal S}} \right.} \right\},\vspace{1.25ex}
\end{equation}
where ${\cal S} \buildrel \Delta \over =\left\{ {0,\frac{{2\pi }}{{{2^{{Q_\phi }}}}}, \ldots ,\frac{{2\pi \left( {{2^{{Q_\phi }}} - 1} \right)}}{{{2^{{Q_\phi }}}}}} \right\}$. In fact, by letting ${Q_\phi } \to \infty $, the model in \eqref{eq:amppha} becomes the continuous RC case.

\subsection{System Model}
\begin{figure}[!htb]
\captionsetup{font={small}}
\begin{center}
  \includegraphics[width=3in,angle=0]{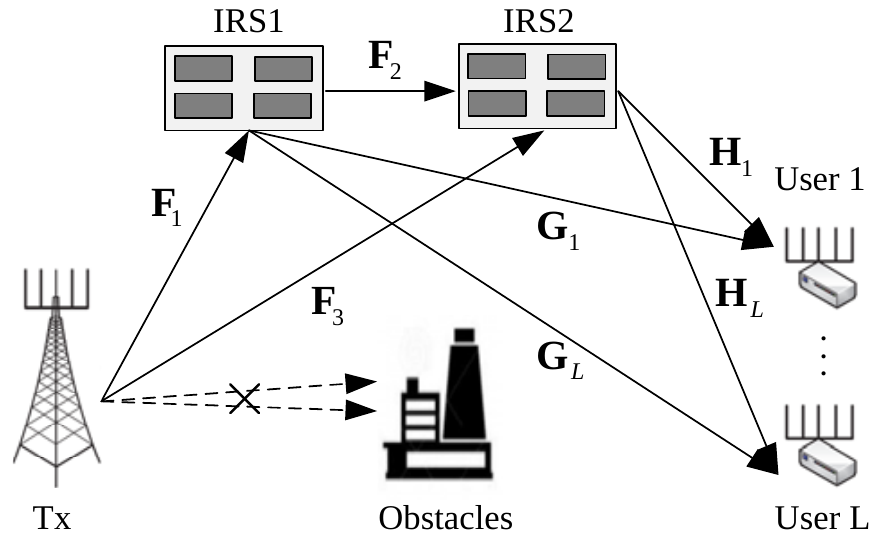}\\
  \caption{The double-IRS-assisted mmWave MIMO system.}
  \label{Fig:model}
\end{center}
\end{figure}

As shown in Fig. \ref{Fig:model}, we consider a double-IRS-assisted network which consists of one Tx, two IRSs, and $L$ users. We assume that Tx and the $l$-th user, $\forall l \in \left\{ {1, \ldots ,L} \right\}$ are equipped with $N_{{\mathrm{TX}}}$ and $N_{U,l}$ antennas, respectively, while the IRSs have $M_1$ and $M_2$ reflection elements, respectively. We denote ${\bf{F}}_1 \in {\mathbb{C}^{M_1 \times {N_{\mathrm{TX}}}}}$, ${{\bf{F}}_2} \in {\mathbb{C}^{{M_2} \times {M_1}}}$, ${{\bf{F}}_3} \in {\mathbb{C}^{{M_2} \times {N_{{\rm{TX}}}}}}$, ${{\bf{G}}_l} \in {\mathbb{C}^{{N_{U,l}} \times {M_1}}}$, and ${{\bf{H}}_l} \in {\mathbb{C}^{{N_{U,l}} \times {M_2}}}$ as the Tx-to-IRS1 link, the IRS1-to-IRS2 link, the Tx-to-IRS2 link, the IRS1-to-$l$-th user link, and the IRS2-to-$l$-th user link, respectively. The Tx-IRS2-IRS1-users link suffers much larger path loss due to the longer link distance, thus is ignored \cite{ZhengTWC2021}. Besides, we assume that the direct link between Tx and users are blocked by obstacles, and all CSI are perfectly obtained at Tx, since we aim to obtain an upper bound of the rate performance. In fact, efficient channel estimation methods for double-IRS-assisted network have been proposed in \cite{YouWCL2021} and \cite{ZhengTC2021}, with On/Off reflection and always-On reflection, respectively. Here, we assume that a hybrid precoding architecture is jointly achieved by Tx and the IRS, where the baseband signal is fully-digitally processed by Tx, and the RF signal send by Tx is reflected by the IRS to obtain analog BF.

Let ${{\bf{s}}_l} \in {\mathbb{C}^{{N_{d,l}} \times 1}}$ be the transmit symbols for the $l$-th user with $N_{d,l}$ being the number of information streams for the $l$-th user. Then, the transmitted signal by Tx is given as ${\bf{x}} = \sum\limits_{l = 1}^L {{{\bf{W}}_l}{{\bf{s}}_l}}$, where ${{\bf{W}}_l} \in {\mathbb{C}^{{N_{\mathrm{TX}}} \times {N_{d,l}}}}$ is the dedicated digital precoding matrix for the $l$-th user. Here, we assume that $\mathbb{E}\left[ {{{\bf{s}}_l}{\bf{s}}_l^H} \right] = {\bf{I}}$ and $\mathbb{E}\left[ {{{\bf{s}}_{l^\prime}}{\bf{s}}_l^H} \right] = {\bf{0}}$ for any $l \ne {l^\prime}$.

Hence, the received signal at the $l$-th user is given as\vspace{1.25ex}
\begin{equation}\label{eq:recsig}
{{\bf{y}}_l} = \left( {{{\bf{H}}_l}{{\bf{\Theta }}_2}{{\bf{F}}_2}{{\bf{\Theta }}_1}{{\bf{F}}_1} + {{\bf{G}}_l}{{\bf{\Theta }}_1}{{\bf{F}}_1} + {{\bf{H}}_l}{{\bf{\Theta }}_2}{{\bf{F}}_3}} \right){\bf{x}} + {{\bf{n}}_l},\vspace{1.25ex}
\end{equation}
where ${{\bf{\Theta }}_1} \in {\mathbb{C}^{{M_1} \times {M_1}}}$ and ${{\bf{\Theta }}_2} \in {\mathbb{C}^{{M_2} \times {M_2}}}$ denote the phase shifters for the two IRSs, respectively, and ${{\bf{n}}_l}$ is the noise vector at the $l$-th user with \({\bf{n}}_l \sim {\cal C}{\cal N}\left( {{\bf{0}}, \sigma_l^2{\bf{I}} } \right)\), where $\sigma_l^2$ is the noise power.

The $l$-th user utilizes the linear decoder matrix ${{\bf{V}}_l} \in {\mathbb{C}^{{N_{d,l}} \times {N_{U,l}}}}$ to obtain an estimation ${{\bf{\hat s}}_l} \in {\mathbb{C}^{{N_{d,l}} \times 1}}$, e.g., ${{\bf{\hat s}}_l} = {{\bf{V}}_l}{{\bf{y}}_l}$. Then, the information rate of the $l$-th user is given as\vspace{1.25ex}
\begin{equation}\label{eq:rate1}
{R_l} = \ln \left| {{\bf{I}} + {{\bf{V}}_l}{{{\bf{\bar H}}}_l}{{\bf{W}}_l}{\bf{W}}_l^H{\bf{\bar H}}_l^H{\bf{V}}_l^H{{\left( {{{\bf{V}}_l}{{\bf{C}}_l}{\bf{V}}_l^H} \right)}^{ - 1}}} \right|,\vspace{1.25ex}
\end{equation}
where ${{{\bf{\bar H}}}_l} = {{\bf{H}}_l}{{\bf{\Theta }}_2}{{\bf{F}}_2}{{\bf{\Theta }}_1}{{\bf{F}}_1} + {{\bf{G}}_l}{{\bf{\Theta }}_1}{{\bf{F}}_1} + {{\bf{H}}_l}{{\bf{\Theta }}_2}{{\bf{F}}_3}$ is the equivalent channel matrix between Tx and the $l$-th user, and ${{\bf{C}}_l} = {{{\bf{\bar H}}}_l}\sum\limits_{j \ne l}^L {{{\bf{W}}_j}{\bf{W}}_j^H{\bf{\bar H}}_l^H}  + \sigma _l^2{\bf{I}}$ is the interference plus noise covariance matrix for the $l$-th user.

Then, by employing the MMSE decoder, the $l$-th decoder matrix is given by{\footnote{In fact, there exist some other decoders for MIMO systems such as the zero forcing (ZF) decoder, the maximum ratio combining (MRC) decoder, the vertical bell labs layered space-times (V-BLAST) decoder, and the maximum likelihood (ML) decoder, etc. According to [1], the MMSE decoder is a kind of linear decoder, which has lower complexity than the non-linear decoders such as the ML decoder and the V-BLAST decoder. In addition, when comparing with other linear decoders, e.g., the ZF decoder and the MRC decoder, since MMSE takes into consideration both interference and noise, the MMSE decoder can obtain better performance than the ZF decoder and the MRC decoder. Thus, to strike a balance between the performance and implementation complexity, we adapt the MMSE decoder in this work.}}
\begin{equation}\label{eq:Vl}
{\bf{V}}_l  = {\bf{W}}_l^H{\bf{\bar H}}_l^H{\left( {\sum\limits_{j = 1}^L {{{{\bf{\bar H}}}_l}{{\bf{W}}_j}{\bf{W}}_j^H{\bf{\bar H}}_l^H + \sigma _l^2{\bf{I}}} } \right)^{ - 1}},
\end{equation}

By substituting \eqref{eq:Vl} into \eqref{eq:rate1}, the achievable information rate for the $l$-th user is\vspace{1.25ex}
\begin{equation}\label{eq:rate}
{R_l} = \ln \left| {{\bf{I}} + {\bf{W}}_l^H{\bf{\bar H}}_l^H{\bf{C}}_l^{ - 1}{{{\bf{\bar H}}}_l}{{\bf{W}}_l}} \right|.\vspace{1.25ex}
\end{equation}
The proof of \eqref{eq:rate} can be found in \cite[Appendix A]{NaghshTSP2019}, we omit the details for brevity.

\subsection{Channel Model}
Due to the small wavelength, mmWave has limited ability to diffract around obstacles. As a result, mmWave channels are usually characterized by the extended Saleh-Valenzuela model \cite{AyachTWC2014}, \cite{AkdenizJSAC2014}, \cite{WangTWC2021}, \cite{Hong2021}.{\footnote{In fact, there exists some other models to describe the mmWave channel such as the 3rd generation partnership project (3GPP) model \cite{3GPP}, \cite{Lu2020}. However, the extended Saleh-Valenzuela model is a perfect fit for modeling the IRS-assisted mmWave channel in our work, especially for high frequency band such as $28\;{\rm{GHz}}$, which is a typical frequency band for implementing mmWave communication. In addition, the channels in IRS-assisted network often have line-of-sight (LoS) and non-line-of-sight (NLoS) components simultaneously. In the extended Saleh-Valenzuela model, both the LoS and NLoS component can be modeled precisely setting the related parameters appropriately. Based on these observations, we adapt the extended Saleh-Valenzuela model in our work.}} For example, ${\bf{F}}_1$ is given by
\begin{equation}\label{eq:svmodel}
{{\bf{F}}_1} = \sum\limits_{q = 1}^{{N_{{\rm{path}}}}} {{\alpha _q}{{\bf{a}}_{\rm{r}}}\left( {\psi _q^{\rm{r}},\beta _q^{\rm{r}}} \right){{\bf{a}}_{\rm{t}}}{{\left( {\psi _q^{\rm{t}},\beta _q^{\rm{t}}} \right)}^H}} ,
\end{equation}
where ${N_{{\rm{path}}}}$ is the number of physical propagation paths in ${\bf{F}}_1$, ${{\alpha _q}}$ is the gain of the $q$-th path in ${\bf{F}}_1$. Here, we assume that ${{\alpha _q}}$ are independently distributed
with ${\cal C}{\cal N}\left( {0,{\kappa ^2}{{10}^{ - 0.1PL\left( D \right)}}} \right)$, $\forall q \in \left\{ {1, \ldots ,{N_{{\rm{path}}}}} \right\}$, where $\kappa = \sqrt {{{{M_1}{N_{{\rm{TX}}}}} \mathord{\left/
 {\vphantom {{{M_1}{N_{{\rm{TX}}}}} {{N_{{\rm{path}}}}}}} \right.
 \kern-\nulldelimiterspace} {{N_{{\rm{path}}}}}}} $ is the normalization factor, and ${PL\left( D \right)}$ is the path loss that depends on the distance $D$ between the two entities associated with ${{\bf{F}}_1}$ \cite{AkdenizJSAC2014}. Besides, the array response vectors associated with the $q$-th path in ${{\bf{F}}_1}$ are respectively denoted by ${{{\bf{a}}_{\rm{r}}}\left( {\psi _q^{\rm{r}},\beta _q^{\rm{r}}} \right)}$ and ${{{\bf{a}}_{\rm{t}}}\left( {\psi _q^{\rm{t}},\beta _q^{\rm{t}}} \right)}$, where $\psi _q^{\rm{r}}\left( {\beta _q^{\rm{r}}} \right)$ and $\psi _q^{\rm{t}}\left( {\beta _q^{\rm{t}}} \right)$ represent the azimuth and (elevation) angles of arrivals and departures (AoAs and AoDs) of the path, respectively.

Here, we assume that uniform planar arrays (UPAs) are employed at all these nodes. Then, the transmit array response vectors ${{{\bf{a}}_{\rm{t}}}{{\left( {\psi _q^{\rm{t}},\beta _q^{\rm{t}}} \right)}}}$ corresponding to the $q$-th path in ${{\bf{F}}_1}$ is given as
\begin{equation}\label{eq:res}
\begin{split}
&{{\bf{a}}_{\rm{t}}}({\psi _q},{\beta _q}) \\
&= \frac{1}{{\sqrt {{N_{{\rm{TX}}}}} }}\left[ {1, \ldots ,{e^{j\frac{{2\pi d}}{\lambda }(m\sin ({\psi _q})\sin ({\beta _q}) + n\cos ({\beta _q}))}},} \right. \\
&~~~~\;\;{\left. { \ldots ,{e^{j\frac{{2\pi d}}{\lambda }((W - 1)\sin (\psi _q )\sin (\beta_q ) + (H - 1)\cos (\beta _q ))}}} \right]^T},
\end{split}
\end{equation}
where $\lambda$ is the signal wavelength, $d$ is the distance between the antennas or IRS elements, $0 \le m < W$ and $0 \le n < H$ denote the horizontal and vertical antenna element indices of the antenna plane, respectively. Besides, the whole antenna array size is ${N_{{\rm{TX}}}} = WH$. Other array response vectors can be similarly defined.

\subsection{Problem Formulation}
Here, we maximize the WSR among these users, by jointly optimizing the digital precoding matrices and the phase shifters, subject to the minimum achievable rate constraints for users. We mainly tackle the continuous RC design, then extend the proposed method to the discrete RC case. Thus, our problem is formulated as:
\begin{subequations}\label{eq:Op}
\begin{align}
&\mathop {\max }\limits_{\left\{ {{{\bf{W}}_l}} \right\}_{l = 1}^L,{\bf{\Theta }}_1, {\bf{\Theta }}_2} \;\;\sum\limits_{l = 1}^L {{\varpi_l}{R_l}}  \label{eq:Opo} \\
&\;\;\;\;\;\;\;\;\;\;\;{\rm{s.t.}}\;\;\sum\limits_{l = 1}^L {{\rm{Tr}}\left( {{{\bf{W}}_l}{\bf{W}}_l^H} \right) \le {P_s}},\label{eq:Opc1} \\
&\;\;\;\;\;\;\;\;\;\;\;\;\;\;\;\;\;\;\left| {{{\left[ {{{\bf{\Theta }}_1}} \right]}_{m,m}}} \right| = 1,\left| {{{\left[ {{{\bf{\Theta }}_2}} \right]}_{m,m}}} \right| = 1,\forall m, \label{eq:Opc2} \\
&\;\;\;\;\;\;\;\;\;\;\;\;\;\;\;\;\;\;{R_l} \ge \Gamma , \label{eq:Opc3}
\end{align}
\end{subequations}
where ${\varpi_l}\left( {0 \le {\varpi_l} \le 1,\sum\limits_{l = 1}^L {{\varpi_l} = 1} } \right)$ is the weight for the $l$-th user, $P_s$ denotes the maximum transmit power, and $\Gamma $ is the minimum rate threshold for these users, respectively.

\section{Joint Transmit Precoding and Phase Shift Design}\label{method}
In this part, we develop a BCD-based method to obtain a suboptimal solution of \eqref{eq:Op}. In particular, we decouple \eqref{eq:Op} into two subproblems, where each subproblem is solved by one efficient method.

\subsection{Problem Transformation}
Here, we transform the non-convex WSR design to a solvable problem by utilizing the MM method. Firstly, by using the determinant identity ${\left| {{\bf{I}} + {\bf{XY}}} \right| = \left| {{\bf{I}} + {\bf{YX}}} \right|}$, we have ${R_l} = \ln \left| {{\bf{I}} + {{{\bf{\bar H}}}_l}{{\bf{W}}_l}{\bf{W}}_l^H{\bf{\bar H}}_l^H{\bf{C}}_l^{ - 1}} \right|$.

In fact, we want to obtain a trackable lower bound for $R_l$. Firstly, we denote ${{\bf{\tilde W}}_l}$, ${\bf{\tilde \Theta }}_1$ and ${\bf{\tilde \Theta }}_2$ as the obtained ${{\bf{W}}_l}$, ${\bf{\Theta }}_1$ and ${\bf{\Theta }}_2$ in the previous iteration, respectively, and introduce auxiliary variables ${{{\bf{\tilde H}}}_l} = {{\bf{H}}_l}{{\bf{\tilde \Theta }}_2}{{\bf{F}}_2}{{\bf{\tilde \Theta }}_1}{{\bf{F}}_1} + {{\bf{G}}_l}{{\bf{\tilde \Theta }}_1}{{\bf{F}}_1} + {{\bf{H}}_l}{{\bf{\tilde \Theta }}_2}{{\bf{F}}_3}$, ${{{\bf{\tilde C}}}_l} = {{{\bf{\tilde H}}}_l}\sum\limits_{j \ne l}^L {{{{\bf{\tilde W}}}_j}{\bf{\tilde W}}_j^H{\bf{\tilde H}}_l^H}  + \sigma _l^2{\bf{I}}$.

Then, according to \cite{NaghshTSP2019}, a lower bound of ${R_l}$ can be established as \vspace{1.25ex}
\begin{equation}\label{eq:rereRl}
{R_l} \ge   \ln \left| {{\bf{E}}{{{\bf{\tilde B}}_l}^{ - 1}}{\bf{E}}} \right| - {\rm{Tr}}\left( {{{{\bf{\tilde A}}}_l}\left( {{\bf{B}}_l - {\bf{\tilde B}}_l} \right)} \right),\vspace{1.25ex}
\end{equation}
where
\begin{subequations}\label{eq:Bl}
\begin{align}
{{\bf{B}}_l} \buildrel \Delta \over = \left[ {\begin{array}{*{20}{c}}
{\bf{I}}&{{\bf{W}}_l^H{\bf{\bar H}}_l^H}\\
{{{{\bf{\bar H}}}_l}{{\bf{W}}_l}}&{{{{\bf{\bar H}}}_l}{{\bf{W}}_l}{\bf{W}}_l^H{\bf{\bar H}}_l^H + {{\bf{C}}_l}}
\end{array}} \right], \\
{{\bf{\tilde B}}_l} \buildrel \Delta \over = \left[ {\begin{array}{*{20}{c}}
{\bf{I}}&{{\bf{\tilde W}}_l^H{\bf{\tilde  H}}_l^H}\\
{{{{\bf{\tilde  H}}}_l}{{\bf{\tilde W}}_l}}&{{{{\bf{\tilde H}}}_l}{{\bf{\tilde W}}_l}{\bf{\tilde W}}_l^H{\bf{\tilde H}}_l^H + {{\bf{C}}_l}}
\end{array}} \right],
\end{align}
\end{subequations}
and ${{\bf{E}} = {{[{\bf{I}}\;\;{\bf{0}}]}^T}}$, ${{\bf{\tilde A}}_l}$ is an auxiliary variable associated with ${{\bf{\tilde B}}_l}$, which is given by ${{\bf{\tilde A}}_l} = {\bf{\tilde B}}_l^{ - 1}{\bf{E}}{\left( {{{\bf{E}}^H}{\bf{\tilde B}}_l^{ - 1}{\bf{E}}} \right)^{ - 1}}{{\bf{E}}^H}{\bf{\tilde B}}_l^{ - 1}$.

Thus, by denoting ${\tilde c}_l = \ln \left| {{{\bf{E}}^H}{{{\bf{\tilde B}}_l}^{ - 1}}{\bf{E}}} \right| + {\rm{Tr}}\left( {{{{\bf{\tilde A}}}_l}{\bf{\tilde B}}_l} \right)$, we have ${R_l} \ge {\tilde c}_l - {\rm{Tr}}\left( {{{{\bf{\tilde A}}}_l}{\bf{B}}_l} \right)$. The MM technique makes use of \eqref{eq:rereRl} and iteratively solves the following problem
\begin{subequations}\label{eq:MM}
\begin{align}
&\mathop {\min }\limits_{\left\{ {{{\bf{W}}_l}} \right\}_{l = 1}^L,{{\bf{\Theta }}_1},{{\bf{\Theta }}_2}} \;\sum\limits_{l = 1}^L {{\varpi _l}{\rm{Tr}}\left( {{{{\bf{\tilde A}}}_l}{{\bf{B}}_l}} \right)} \label{eq:MMo} \\
&\;\;\;\;\;\;\;\;\;\;{\rm{s.t.}}\;\;\;\;\;\;\;\;\;\;\;\eqref{eq:Opc1},\eqref{eq:Opc2},\label{eq:MMc1} \\
&\;\ln \left| {{\bf{E}}{{{\bf{\tilde B}}}_l}^{ - 1}{\bf{E}}} \right| - {\rm{Tr}}\left( {{{{\bf{\tilde A}}}_l}\left( {{{\bf{B}}_l} - {{{\bf{\tilde B}}}_l}} \right)} \right) \ge \Gamma. \label{eq:MMc2}
\end{align}
\end{subequations}

Moreover, according to \cite{NaghshTSP2019}, ${{\bf{\tilde A}}_l}$ can be decomposed as\vspace{1.25ex}
\begin{equation}\label{eq:Al}
{{\bf{\tilde A}}_l} = \left[ {\begin{array}{*{20}{c}}
{{\bf{A}}_l^{11}}&{{\bf{A}}_l^{12}}\\
{{\bf{A}}_l^{21}}&{{\bf{A}}_l^{22}}
\end{array}} \right] = \left[ {\begin{array}{*{20}{c}}
{{\bf{\tilde X}}_l^H}&{{{\bf{\tilde Y}}_l}}\\
{{\bf{\tilde Y}}_l^H}&{{\bf{\tilde Y}}_l^H{\bf{\tilde X}}_l^{ - 1}{{\bf{\tilde Y}}_l}}
\end{array}} \right],\vspace{1.25ex}
\end{equation}
where ${\bf{\tilde  X}}_l$, ${\bf{\tilde  Y}}_l$ are respectively, given by ${{\bf{\tilde X}}_l} = {\bf{I}} + {\bf{\tilde W}}_l^H{\bf{\tilde H}}_l^H{\bf{\tilde C}}_l^{ - 1}{{\bf{\tilde H}}_l}{{\bf{\tilde W}}_l}$, and
${{\bf{\tilde Y}}_l} =  - {\bf{\tilde W}}_l^H{\bf{\tilde H}}_l^H{\bf{\tilde C}}_l^{ - 1}$.

Thus, the following equation can be obtained\vspace{1.25ex}
\begin{equation}\label{eq:invAB}
\begin{split}
&{{{\bf{\tilde A}}}_l}{{\bf{B}}_l} = {\bf{A}}_l^{11} + 2\Re \left\{ {{\bf{A}}_l^{12}{{{\bf{\bar H}}}_l}{{\bf{W}}_l}} \right\} + \\
&\sigma _l^2{\bf{A}}_l^{22} +{\bf{A}}_l^{22}{{{\bf{\bar H}}}_l}\left( {\sum\limits_{l = 1}^L {{{\bf{W}}_l}{\bf{W}}_l^H} } \right){\bf{\bar H}}_l^H.\vspace{1.25ex}
\end{split}
\end{equation}

Via the above procedure, we transform \eqref{eq:MM} into the following equivalent problem
\begin{subequations}\label{eq:Ap}
\begin{align}
\begin{split}
&\mathop {\min }\limits_{\left\{ {{{\bf{W}}_l}} \right\}_{l = 1}^L,{{\bf{\Theta }}_1},{{\bf{\Theta }}_2}} \;\;\sum\limits_{l = 1}^L {2{\varpi_l}\Re \left\{ {{\rm{Tr}}\left( {{\bf{A}}_l^{12}{{{\bf{\bar H}}}_l}{{\bf{W}}_l}} \right)} \right\}} \\
& + \sum\limits_{l = 1}^L {{\varpi_l}} {\rm{Tr}}\left( {{\bf{A}}_l^{22}{{{\bf{\bar H}}}_l}\left( {\sum\limits_{l = 1}^L {{{\bf{W}}_l}{\bf{W}}_l^H} } \right){\bf{\bar H}}_l^H} \right) \label{eq:Apo}\end{split}  \\
&\;\;\;\;\;\;\;\;\;\;{\rm{s.t.}}\;\;\;\;\; \eqref{eq:Opc1},\eqref{eq:Opc2}, \label{eq:Apc1} \\
\begin{split}
&\;{\rm{Tr}}\left( {{\bf{A}}_l^{22}{{{\bf{\bar H}}}_l}\left( {\sum\limits_{l = 1}^L {{{\bf{W}}_l}{\bf{W}}_l^H} } \right){\bf{\bar H}}_l^H} \right) \\
&\; + 2\Re \left\{ {{\rm{Tr}}\left( {{\bf{A}}_l^{12}{{{\bf{\bar H}}}_l}{{\bf{W}}_l}} \right)} \right\} \le  - \Gamma  + \ln \left| {{\bf{E}}{{{\bf{\tilde B}}}_l}^{ - 1}{\bf{E}}} \right| \\
& \;+ {\rm{ Tr}}\left( {{{{\bf{\tilde A}}}_l}{{{\bf{\tilde B}}}_l}} \right) - {\bf{A}}_l^{11} - \sigma _l^2{\bf{A}}_l^{22}, \forall l. \label{eq:Apc2} \end{split}
\end{align}
\end{subequations}

The main advantages of \eqref{eq:Ap} is that the objective function is in a quadratic form, which is convex with respect to (w.r.t.) a given variable when fixing the others. Then, we propose a BCD method to obtain a suboptimal solution of \eqref{eq:Ap}.

\subsection{Digital Precoding Optimization}
Firstly, we solve the subproblem w.r.t. $\left\{ {{{\bf{W}}_l}} \right\}_{l = 1}^L$ with fixed ${\bf{\Theta }}_1$ and ${\bf{\Theta }}_2$. By denoting ${{\bf{w}}_l} = {\rm{vec}}\left( {{{\bf{W}}_l}} \right)$ and utilizing the equalities ${\rm{Tr}}\left( {{\bf{XY}}} \right) = {\rm{vec}}{\left( {{{\bf{X}}^H}} \right)^H}{\rm{vec}}\left( {\bf{Y}} \right)$ and ${\rm{Tr}}\left( {{{\bf{X}}^H}{\bf{YXZ}}} \right) = {\rm{vec}}{\left( {\bf{X}} \right)^H}\left( {{{\bf{Z}}^T} \otimes {\bf{Y}}} \right){\rm{vec}}\left( {\bf{X}} \right)$, \eqref{eq:Ap} can be rewritten as
\begin{subequations}\label{eq:reRp}
\begin{align}
&\mathop {\min }\limits_{\left\{ {{{\bf{w}}_l}} \right\}_{l = 1}^L} \;\;\sum\limits_{l = 1}^L {{\varpi_l}\left( {{\bf{w}}_l^H{{\bf{V}}_l}{{\bf{w}}_l} + 2\Re \left\{ {{\bf{v}}_l^H{{\bf{w}}_l}} \right\}} \right)}  \label{eq:reRpo} \\
&\;\;\;\;\;{\rm{s.t.}}\;\;\sum\limits_{l = 1}^L {{\bf{w}}_l^H{{\bf{w}}_l} \le {P_s}}, \label{eq:reRpc1} \\
\begin{split}
&{\bf{w}}_l^H{{\bf{V}}_l}{{\bf{w}}_l} + 2\Re \left\{ {{{\bf{v}}_l^H}{{\bf{w}}_l}} \right\} \le  - \Gamma  + \ln \left| {{\bf{E}}{{{\bf{\tilde B}}}_l}^{ - 1}{\bf{E}}} \right|\\
& + {\rm{Tr}}\left( {{{{\bf{\tilde A}}}_l}{{{\bf{\tilde B}}}_l}} \right) - {\bf{A}}_l^{11} - \sigma _l^2{\bf{A}}_l^{22}, \forall l, \label{eq:reRpc2}
\end{split}
\end{align}
\end{subequations}
where ${{\bf{V}}_l} = {\bf{I}} \otimes \left( {{\bf{\bar H}}_l^H{\bf{A}}_l^{22}{{{\bf{\bar H}}}_l}} \right)$, and ${{\bf{v}}_l} = {\rm{vec}}\left( {{{\left( {{\bf{A}}_l^{12}{{{\bf{\bar H}}}_l}} \right)}^H}} \right)$, respectively.

Note that \eqref{eq:reRp} is a QCQP problem, which can be efficiently solved by the optimization toolbox CVX \cite{GrantCVX}.

\subsection{Phase Shifter Optimization}
Here, we handle the phase shifter optimization. Since ${\bf{\Theta }}_1$ and ${\bf{\Theta }}_2$ is symmetric in ${{\bf{\bar H}}_l}$ and ${{\bf{\tilde H}}_l}$, the proposed method for ${\bf{\Theta }}_1$ is suitable for ${\bf{\Theta }}_2$, vice versa. Thus, in the following, we focus on the design of ${\bf{\Theta }}_1$ with given $\left\{ {{{\bf{W}}_l}} \right\}_{l = 1}^L$ and ${\bf{\Theta }}_2$. Specifically, by denoting ${\bf{\Xi }} = \sum\limits_{l = 1}^L {{{\bf{W}}_l}{\bf{W}}_l^H}$, and substituting ${{{\bf{\bar H}}}_l} = {{\bf{H}}_l}{{\bf{\Theta }}_2}{{\bf{F}}_2}{{\bf{\Theta }}_1}{{\bf{F}}_1} + {{\bf{G}}_l}{{\bf{\Theta }}_1}{{\bf{F}}_1} + {{\bf{H}}_l}{{\bf{\Theta }}_2}{{\bf{F}}_3}$ into the objective function and neglecting the irrelevant terms, we obtain the following problem
\begin{subequations}\label{eq:reforpro}
\begin{align}
\begin{split}
&\mathop {\min }\limits_{{{\bf{\Theta }}_1}} \;\sum\limits_{l = 1}^L {{\varpi _l}} {\rm{Tr}}\left( {{\bf{A}}_l^{22}{\bf{\Psi \Xi }}{{\bf{\Psi }}^H}} \right) \\
&~~~~~~+ \sum\limits_{l = 1}^L {2{\varpi _l}\Re \left\{ {{\rm{Tr}}\left( {{\bf{A}}_l^{12}{\bf{\Psi }}{{\bf{W}}_l}} \right)} \right\}} \\
&~~~~~~ + \sum\limits_{l = 1}^L {2{\varpi _l}} \Re \left\{ {{\rm{Tr}}\left( {{\bf{A}}_l^{22}{\bf{\Psi \Xi F}}_3^H{\bf{\Theta }}_2^H{\bf{H}}_l^H} \right)} \right\}  \end{split}  \label{eq:reforproo} \\
&~~{\rm{s.t.}}\;\;\;\left| {{{\left[ {{{\bf{\Theta }}_1}} \right]}_{m,m}}} \right| = 1, \forall m, \label{eq:reforproc1} \\
\begin{split}
&{\rm{Tr}}\left( {{\bf{A}}_l^{22}{\bf{\Psi \Xi }}{{\bf{\Psi }}^H}} \right) + 2\Re \left\{ {{\rm{Tr}}\left( {{\bf{A}}_l^{12}{\bf{\Psi }}{{\bf{W}}_l}} \right)} \right\} \\
& + 2\Re \left\{ {{\rm{Tr}}\left( {{\bf{A}}_l^{22}{\bf{\Psi \Xi F}}_3^H{\bf{\Theta }}_2^H{\bf{H}}_l^H} \right)} \right\} \le - {{\tilde \Gamma }_l}, \forall l, \end{split}  \label{eq:reforproc2}
\end{align}
\end{subequations}
where ${\bf{\Psi }} = \left( {{{\bf{H}}_l}{{\bf{\Theta }}_2}{{\bf{F}}_2} + {{\bf{G}}_l}} \right){{\bf{\Theta }}_1}{{\bf{F}}_1}$, and $- {{\tilde \Gamma }_l} = - {\Gamma } + \ln \left| {{\bf{E}}{{{\bf{\tilde B}}}_l}^{ - 1}{\bf{E}}} \right| + {\rm{Tr}}\left( {{{{\bf{\tilde A}}}_l}{{{\bf{\tilde B}}}_l}} \right) - {\bf{A}}_l^{11} - \sigma _l^2{\bf{A}}_l^{22}$, respectively.

To solve \eqref{eq:reforpro}, we introduce the following Lemma.

{\textit{Lemma 1 \cite{PanJSAC2020}:}} Let ${\bf{C}}_1 \in \mathbb{C}^{m \times m}$, ${\bf{C}}_2 \in \mathbb{C}^{m \times m}$. Then, for a diagonal matrix ${\bf{E}}={\rm{Diag}}\left(e_{1}, \ldots, e_{m}\right)$, and ${\bf{e}} = {\rm{diag}}\left( {\bf{E}} \right)$, the following relationships hold:\vspace{1.25ex}
\begin{equation*}\label{eq:Lemma}
\begin{split}
&{\mathop{\rm Tr}\nolimits} \left( {{{\bf{E}}^H}{{\bf{C}}_1}{\bf{E}}{{\bf{C}}_2}} \right) = {{\bf{e}}^H}\left( {{{\bf{C}}_1} \odot {\bf{C}}_2^T} \right){\bf{e}}, \vspace{1.25ex}\\
&\operatorname{Tr}\left(\mathbf{EC}_{2}\right)=\mathbf{1}^{T}\left(\mathbf{E} \odot \mathbf{C}_{2}^{T}\right) \mathbf{1}=\mathbf{e}^{T} \mathbf{c}_{2}, \vspace{1.25ex}\\
&\operatorname{Tr}\left(\mathbf{E}^{H} \mathbf{C}_{2}^{H}\right)=\mathbf{c}_{2}^{H} \mathbf{e}^*,\vspace{1.25ex}
\end{split}
\end{equation*}
where ${{\bf{c}}_2} = {\rm{diag}}\left( {{{\bf{C}}_2}} \right)$.

By utilizing Lemma 1, we have\vspace{1.25ex}
\begin{subequations}\label{eq:relation1}
\begin{align}
&2\Re \left\{ {{\rm{Tr}}\left( {{\bf{A}}_l^{12}{\bf{\Psi }}{{\bf{W}}_l}} \right)} \right\}= 2\Re \left\{ {{\bf{q}}_{1,l}^T{{\boldsymbol{\theta }}_1}} \right\}, \vspace{1.25ex}\\
\begin{split}
&{\rm{Tr}}\left( {2\Re \left\{ {{\bf{A}}_l^{22}{{\bf{H}}_l}{{\bf{\Theta }}_2}{{\bf{F}}_2}{{\bf{\Theta }}_1}{{\bf{F}}_1}{\bf{\Xi}} {\bf{F}}_3^H{\bf{\Theta }}_2^H{\bf{H}}_l^H} \right\}} \right) \vspace{1.25ex}\\
&=2\Re \left\{ {{\bf{q}}_{2,l}^T{{\boldsymbol{\theta }}_1}} \right\}, \end{split}\vspace{1.25ex}\\
\begin{split}
&{\rm{Tr}}\left( {2\Re \left\{ {{\bf{A}}_l^{22}{{\bf{G}}_l}{{\bf{\Theta }}_1}{{\bf{F}}_1}{\bf{\Xi}} {\bf{F}}_3^H{\bf{\Theta }}_2^H{\bf{H}}_l^H} \right\}} \right) \vspace{1.25ex}\\
&= 2\Re \left\{ {{\bf{q}}_{3,l}^T{{\boldsymbol{\theta }}_1}} \right\}, \vspace{1.25ex}\end{split}\\
\begin{split}
&{\rm{Tr}}\left( {{\bf{\Theta }}_1^H{\bf{F}}_2^H{\bf{\Theta }}_2^H{\bf{H}}_l^H{\bf{A}}_l^{22}{{\bf{H}}_l}{{\bf{\Theta }}_2}{{\bf{F}}_2}{{\bf{\Theta }}_1}{{\bf{F}}_1}{\bf{\Xi F}}_1^H} \right),\vspace{1.25ex}\\
&+ {\rm{Tr}}\left( {{\bf{\Theta }}_1^H{\bf{G}}_l^H{\bf{A}}_l^{22}{{\bf{G}}_l}{{\bf{\Theta }}_1}{{\bf{F}}_1}{\bf{\Xi F}}_1^H} \right) = {\boldsymbol{\theta }}_1^H{{\bf{U}}_{1,l}}{{\boldsymbol{\theta }}_1}, \vspace{1.25ex}\end{split}\\
&{\rm{Tr}}\left( {{\bf{\Theta }}_1^H{\bf{F}}_2^H{\bf{\Theta }}_2^H{\bf{H}}_l^H{\bf{A}}_l^{22}{{\bf{G}}_l}{{\bf{\Theta }}_1}{{\bf{F}}_1}{\bf{\Xi}}{\bf{F}}_1^H} \right) = {\boldsymbol{\theta }}_1^H{{\bf{U}}_{2,l}}{{\boldsymbol{\theta }}_1}, \vspace{1.25ex}\\
&{\rm{Tr}}\left( {{{\bf{F}}_1}{\bf{\Xi}}{\bf{F}}_1^H{\bf{\Theta }}_1^H{\bf{G}}_l^H{\bf{A}}_l^{22}{{\bf{H}}_l}{{\bf{\Theta }}_2}{{\bf{F}}_2}{{\bf{\Theta }}_1}} \right) = {\boldsymbol{\theta }}_1^H{{\bf{U}}_{3,l}}{{\boldsymbol{\theta }}_1},\vspace{1.25ex}
\end{align}
\end{subequations}
where the related parameters are, respectively, given by\vspace{1.25ex}
\begin{subequations}\label{eq:relation2}
\begin{align}
&{{\bf{q}}_{1,l}} = {\rm{diag}}\left( {{{\bf{F}}_1}{{\bf{W}}_l}{\bf{A}}_l^{12}{{\bf{H}}_l}{{\bf{\Theta }}_2}{{\bf{F}}_2} + {{\bf{F}}_1}{{\bf{W}}_l}{\bf{A}}_l^{12}{{\bf{G}}_l}} \right),\vspace{1.25ex}\\
&{{\bf{q}}_{2,l}} = {\rm{diag}}\left( {{{\bf{F}}_1}{\bf{\Xi F}}_3^H{\bf{\Theta }}_2^H{\bf{H}}_l^H{\bf{A}}_l^{22}{{\bf{H}}_l}{{\bf{\Theta }}_2}{{\bf{F}}_2}} \right), \vspace{1.25ex}\\
&{{\bf{q}}_{3,l}} = {\rm{diag}}\left( {{{\bf{F}}_1}{\bf{\Xi F}}_3^H{\bf{\Theta }}_2^H{\bf{H}}_l^H{\bf{A}}_l^{22}{{\bf{G}}_l}} \right), \vspace{1.25ex}\\
\begin{split}
&{{\bf{U}}_{1,l}} = \left( {{\bf{F}}_2^H{\bf{\Theta }}_2^H{\bf{H}}_l^H{\bf{A}}_l^{22}{{\bf{H}}_l}{{\bf{\Theta }}_2}{{\bf{F}}_2} + {\bf{G}}_l^H{\bf{A}}_l^{22}{{\bf{G}}_l}} \right)\vspace{1.25ex}\\
& \odot {\left( {{{\bf{F}}_1}{\bf{\Xi}}{\bf{F}}_1^H} \right)^T}, \vspace{1.25ex}\end{split} \\
&{{\bf{U}}_{2,l}} = \left( {{\bf{F}}_2^H{\bf{\Theta }}_2^H{\bf{H}}_l^H{\bf{A}}_l^{22}{{\bf{G}}_l}} \right) \odot {\left( {{{\bf{F}}_1}{\bf{\Xi}}{\bf{F}}_1^H} \right)^T},\vspace{1.25ex} \\
&{{\bf{U}}_{3,l}} = \left( {{\bf{G}}_l^H{\bf{A}}_l^{22}{{\bf{H}}_l}{{\bf{\Theta }}_2}{{\bf{F}}_2}} \right) \odot {\left( {{{\bf{F}}_1}{\bf{\Xi}}{\bf{F}}_1^H} \right)^T}.\vspace{1.25ex}
\end{align}
\end{subequations}

Based on these equations, we obtain the following problem w.r.t. ${\boldsymbol{\theta }}_1$
\begin{subequations}\label{eq:rephase0}
\begin{align}
&\mathop {\min }\limits_{{{\boldsymbol{\theta }}_1}} \;\;\sum\limits_{l = 1}^L {{\varpi_l}} \left( {{\boldsymbol{\theta }}_1^H{{\bf{U}}_l}{{\boldsymbol{\theta }}_1} + 2\Re \left\{ {{\boldsymbol{\theta }}_1^H{\bf{q}}_l} \right\}} \right) \label{eq:rephaseo0} \\
&\;\;{\rm{s.t.}}\;\;\left| {{{\left[ {{{\boldsymbol{\theta }}_1}} \right]}_m}} \right| = 1,\forall m, \label{eq:rephasec10} \\
&\;\;\;\;\;\;\;\;\;\;{\boldsymbol{\theta }}_1^H{{\bf{U}}_l}{{\boldsymbol{\theta }}_1} + 2\Re \left\{ {{\boldsymbol{\theta }}_1^H{{\bf{q}}_l}} \right\} \le {{\tilde \Gamma }_l}, \forall l,
\label{eq:rephasec20}
\end{align}
\end{subequations}
where ${{\bf{U}}_l} = {{\bf{U}}_{1,l}} + {{\bf{U}}_{2,l}} + {{\bf{U}}_{3,l}}$, and ${{\bf{q}}_l} = {\bf{q}}_{1,l}^* + {\bf{q}}_{2,l}^* + {\bf{q}}_{3,l}^* $, respectively.

\begin{figure*}[!htb]
\captionsetup{font={small}}
\hspace{0.1in}
\subfigure[Tangent space and Riemannian gradient.]{
\label{fig:RMO:a} 
\includegraphics[width=2.1in]{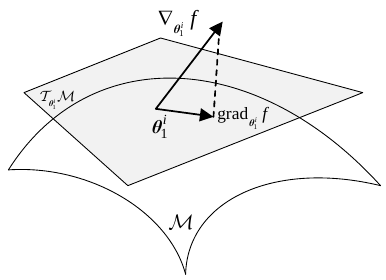}}
\hspace{0.1in}
\subfigure[Vector transport.]{
\label{fig:RMO:b} 
\includegraphics[width=2.1in]{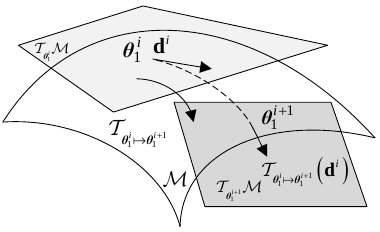}}
\hspace{0.1in}
\subfigure[Retraction.]{
\label{fig:RMO:c} 
\includegraphics[width=2.1in]{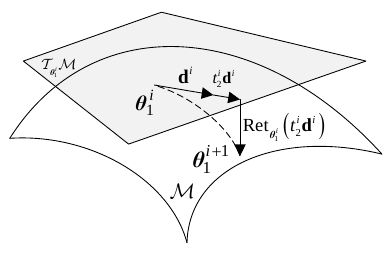}}
\caption{An example of the key steps in manifold optimization.}
\label{fig:RMO} 
\end{figure*}

To handle the QoS constraint \eqref{eq:rephasec20}, a price mechanism is introduced to solve \eqref{eq:rephase0}.
Firstly, we need to convexity the non-smooth constraints \eqref{eq:rephasec20} by a smooth approximation using the log-sum-exp inequality (see e.g., \cite{Xu2001}), i.e.,\vspace{1.25ex}
\begin{equation}\label{eq:Lemma1}
\mathop {\max }\limits_{\forall l \in \left\{ {1, \ldots ,L} \right\}} \;{x_l} \le \frac{1}{\mu }\ln \left( {\sum\limits_{l = 1}^L {{e^{\mu {x_l}}}} } \right) \le \mathop {\max }\limits_{\forall l \in \left\{ {1, \ldots ,L} \right\}} \;{x_l} + \frac{1}{\mu }\ln L,
\end{equation}
for ${x_l} \in \mathbb{R},\forall l$, and $\mu  > 0$ is the smoothing parameter. By applying \eqref{eq:Lemma1} to \eqref{eq:rephasec20}, we obtain a smooth constraint as \vspace{1.25ex}
\begin{equation}\label{eq:Lemma2}
\frac{1}{\mu }\ln \left( {\sum\limits_{l = 1}^L {{e^{\mu \left( {{\boldsymbol{\theta }}_1^H{{\bf{U}}_l}{{\boldsymbol{\theta }}_1} + 2\Re \left\{ {{\boldsymbol{\theta }}_1^H{{\bf{q}}_l}} \right\}} \right)}}} } \right) \le {{\tilde \Gamma }_l}.
\end{equation}
Thus, \eqref{eq:rephase0} can be approximated as
\begin{subequations}\label{eq:arephase0}
\begin{align}
&\mathop {\min }\limits_{{{\boldsymbol{\theta }}_1}} \;\;\sum\limits_{l = 1}^L {{\varpi_l}} \left( {{\boldsymbol{\theta }}_1^H{{\bf{U}}_l}{{\boldsymbol{\theta }}_1} + 2\Re \left\{ {{\boldsymbol{\theta }}_1^H{\bf{q}}_l} \right\}} \right) \label{eq:arephaseo0} \\
&\;\;{\rm{s.t.}}\;\;\left| {{{\left[ {{{\boldsymbol{\theta }}_1}} \right]}_m}} \right| = 1,\forall m, \label{eq:arephasec10} \\
&\;\;\;\;\;\;\;\;\;\;\sum\limits_{l = 1}^L {{e^{\mu \left( {{\boldsymbol{\theta }}_1^H{{\bf{U}}_l}{{\boldsymbol{\theta }}_1} + 2\Re \left\{ {{\boldsymbol{\theta }}_1^H{{\bf{q}}_l}} \right\}} \right)}}}  \le {e^{\mu {{\tilde \Gamma }_l}}}.
\label{eq:arephasec20}
\end{align}
\end{subequations}
Then, by introducing a non-negative price $\rho$ and adding \eqref{eq:arephasec20} into \eqref{eq:arephaseo0}, we obtain the following problem
\begin{subequations}\label{eq:rephase}
\begin{align}
\begin{split}
&\mathop {\min }\limits_{{{\boldsymbol{\theta }}_1}} {\cal L}\left( {{{\boldsymbol{\theta }}_1},\rho } \right) = \sum\limits_{l = 1}^L {{\varpi _l}\left( {{\boldsymbol{\theta }}_1^H{{\bf{U}}_l}{{\boldsymbol{\theta }}_1} + 2\Re \left\{ {{\boldsymbol{\theta }}_1^H{{\bf{q}}_l}} \right\}} \right)} \\
&~~~~~~~~~~+ \rho \left( {\sum\limits_{l = 1}^L {{e^{\mu\left( {{\boldsymbol{\theta }}_1^H{{\bf{U}}_l}{{\boldsymbol{\theta }}_1} + 2\Re \left\{ {{\boldsymbol{\theta }}_1^H{{\bf{q}}_l}} \right\}} \right)}}}  - {e^{\mu {{\tilde \Gamma }_l}}}} \right) \end{split} \label{eq:rephaseo} \\
&\;\;{\rm{s.t.}}\;\;\left| {{{\left[ {{{\boldsymbol{\theta }}_1}} \right]}_m}} \right| = 1,\forall m. \label{eq:rephasec1}
\end{align}
\end{subequations}

Next, we focus on obtaining the optimal ${\boldsymbol{\theta }}_1$ with a given $\rho$. In fact, the objective function \eqref{eq:rephaseo} is a quadratic form, while the main difficulty is the UMC \eqref{eq:rephasec1}, which constitutes a Riemannian manifold \cite{AlhujailiTSP2019}. In fact, the search space of \eqref{eq:rephase} is the product of complex manifold ${\cal M} = \left\{ {\left. {{{\boldsymbol{\theta }}_1} \in {\mathbb{C}^{{M_1}\times 1}}} \right|\left| {{{\left[ {{{\boldsymbol{\theta }}_1}} \right]}_1}} \right| =  \ldots \left| {{{\left[ {{{\boldsymbol{\theta }}_1}} \right]}_{{M_1}}}} \right| = 1} \right\}$, which is a Riemannian manifold of ${\mathbb{C}^{{M_1}\times 1}}$. Then, the following three main steps are needed in each iteration to obtain ${{\boldsymbol{\theta}}_1}$.

Firstly, for any point ${{\boldsymbol{\theta}}_1}$, the tangent space is given by \vspace{1.25ex}
\begin{equation}\label{eq:manifold}
{{\cal T}_{{\boldsymbol{\theta }}_1^i}}{\cal M} = \left\{ {\left. {{\bf{z}} \in {\mathbb{C}^{{M_1}}}} \right|{{\left[ {{\bf{z}}{{\left( {{\boldsymbol{\theta }}_1^i} \right)}^H}} \right]}_{m,m}} = {\bf{0}},\forall m \in {\cal M}} \right\},\vspace{1.25ex}
\end{equation}
where ${\bf{z}}$ is the tangent vector at ${{\boldsymbol{\theta}}_1^i}$. The Riemannian gradient ${\rm{gra}}{{\rm{d}}_{{\boldsymbol{\theta}}_1^i}}f$ at ${{\boldsymbol{\theta}}_1^i}$ is the tangent vector which leads to the steepest decrease of the objective function, and given by\vspace{1.25ex}
\begin{equation}\label{eq:Rieg}
{\mathrm{gra}}{{\mathrm{d}}_{{\boldsymbol{\theta}}_1^i}}f = {\nabla _{{\boldsymbol{\theta}}_1^i}}f - \Re \left\{ {{\nabla _{{{\boldsymbol{\theta}}_1^i}}}f \odot {{\left( {{{\boldsymbol{\theta}}_1^i}} \right)}^*}} \right\} \odot {{\boldsymbol{\theta}}_1^i},\vspace{1.25ex}
\end{equation}
where the Euclidean gradient ${\nabla _{{{{\boldsymbol{\theta}}_1^i}}}}f$ of \eqref{eq:rephase} is\vspace{1.25ex}
\begin{equation}\label{eq:Eieg}
{\nabla _{{{{\boldsymbol{\theta}}_1^i}}}}f= 2{\bf{U }}{{{\boldsymbol{\theta}}_1^i}} +2 {\bf{q }^* },\vspace{1.25ex}
\end{equation}
with ${\bf{U}} = \sum\nolimits_{l = 1}^L {\left( {{\varpi _l} + \rho \sum\limits_{l = 1}^L {{e^{\mu \left( {{{\left( {{\boldsymbol{\theta }}_1^i} \right)}^H}{{\bf{U}}_l}{\boldsymbol{\theta }}_1^i + 2\Re \left\{ {{{\left( {{\boldsymbol{\theta }}_1^i} \right)}^H}{{\bf{q}}_l}} \right\}} \right)}}} } \right)} {{\bf{U}}_l}$, and ${\bf{q}} = \sum\nolimits_{l = 1}^L {\left( {{\varpi _l} + \rho \sum\limits_{l = 1}^L {{e^{\mu \left( {{{\left( {{\boldsymbol{\theta }}_1^i} \right)}^H}{{\bf{U}}_l}{\boldsymbol{\theta }}_1^i + 2\Re \left\{ {{{\left( {{\boldsymbol{\theta }}_1^i} \right)}^H}{{\bf{q}}_l}} \right\}} \right)}}} } \right){{\bf{q}}_l}} $, respectively. The concept of the tangent space and Riemannian gradient is shown in Fig. \ref{fig:RMO:a}.

Secondly, the conjugate gradient for the Riemannian manifold is updated by\vspace{1.25ex}
\begin{equation}\label{eq:tang}
{{\bf{d}}^{i + 1}} =  - {\nabla _{{{\boldsymbol{\theta}}_1^{i+1}}}}f + {t_1^{i+1}}{{\bf{d}}^i},\vspace{1.25ex}
\end{equation}
where ${\bf{d}}^i$ is the search direction at ${{\boldsymbol{\theta}}_1^{i+1}}$, and $t_1^{i+1}$ is the Polak-Ribiere parameter to obtain fast convergence \cite{Absil2009}, which is given by
\begin{equation}\label{eq:Polak}
t_1^{i+1} = \frac{{\Re \left\{ {\nabla _{{{\boldsymbol{\theta}}_1^{i+1}}}^Hf\left( {{\nabla _{{{\boldsymbol{\theta}}_1^{i+1}}}}f - {\nabla _{{{\boldsymbol{\theta}}_1^i}}}f} \right)} \right\}}}{{\nabla _{{{\boldsymbol{\theta}}_1^i}}^Hf{\nabla _{{{\boldsymbol{\theta}}_1^i}}}f}}.
\end{equation}

However, ${\bf{d}}^i$ and ${\bf{d}}^{i + 1}$ in \eqref{eq:tang} exist in two different spaces
${{\cal T}_{{{\boldsymbol{\theta}}_1^i}}}{\cal M}$ and ${{\cal T}_{{{\boldsymbol{\theta}}_1^{i+1}}}}{\cal M}$, thus the search direction cannot be directly obtained. To handle this problem, a transport operation which maps ${\bf{d}}^i$ to ${{\cal T}_{{{\boldsymbol{\theta}}_1^{i+1}}}}{\cal M}$ is proposed, and given by\vspace{1.25ex}
\begin{equation}\label{eq:transport}
\begin{split}
&{{\cal T}_{{{\boldsymbol{\theta}}_1^i} \mapsto {{\boldsymbol{\theta}}_1^{i+1}}}}\left( {{{\bf{d}}^i}} \right) \buildrel \Delta \over = {{\cal T}_{{{\boldsymbol{\theta}}_1^i}}}{\cal M} \mapsto {{\cal T}_{{{\boldsymbol{\theta}}_1^{i+1}}}}{\cal M},\vspace{1.25ex} \\
&{{\bf{d}}^i} \mapsto {{\bf{d}}^i} - \Re \left\{ {{{\bf{d}}_i} \odot {{\left( {{{\boldsymbol{\theta}}_1^{i+1}}} \right)}^*}} \right\} \odot {{\boldsymbol{\theta}}_1^{i+1}}.\vspace{1.25ex}
\end{split}
\end{equation}

Similarly to \eqref{eq:tang}, the update method for the search direction on ${\cal M}$ is given by\vspace{1.25ex}
\begin{equation}\label{eq:cgd}
{{\bf{d}}^{i + 1}} =  - {\mathrm{gra}}{{\mathrm{d}}_{{{\boldsymbol{\theta}}_1^{i+1}}}}f + t_1^i{{\cal T}_{{{\boldsymbol{\theta}}_1^i} \mapsto {{\boldsymbol{\theta}}_1^{i+1}}}}\left( {{{\bf{d}}^i}} \right).\vspace{1.25ex}
\end{equation}
The vector transport operation is shown in Fig. \ref{fig:RMO:b}.

Thirdly, after obtaining ${\bf{d}}^i$ at ${{\boldsymbol{\theta}}_1^i}$, a retraction step is utilized to make the obtained point remains on the manifold. To be specific, the retraction operation for ${\bf{d}}^i$ at ${{\boldsymbol{\theta}}_1^i}$ is given by
\begin{equation}\label{eq:Proj}
\begin{split}
&{{\mathop{\rm {Ret}}\nolimits} _{{\boldsymbol{\theta}}_1^i}}\left( {t_2^i{{\bf{d}}^i}} \right) \buildrel \Delta \over = {{\cal T}_{{{\boldsymbol{\theta}}_1^i}}}{\cal M} \mapsto {\cal M}:\\
&{\left[ {t_2^i{{\bf{d}}^i}} \right]_m} \mapsto {{{{\left[ {{{\boldsymbol{\theta}}_1^i} + t_2^i{{\bf{d}}^i}} \right]}_m}} \mathord{\left/
 {\vphantom {{{{\left[ {{{\boldsymbol{\theta}}_1^i} + t_2^i{{\bf{d}}^i}} \right]}_m}} {\left| {{{\left[ {{{\boldsymbol{\theta}}_1^i} + t_2^i{{\bf{d}}^i}} \right]}_m}} \right|}}} \right.
 \kern-\nulldelimiterspace} {\left| {{{\left[ {{{\boldsymbol{\theta}}_1^i} + t_2^i{{\bf{d}}^i}} \right]}_m}} \right|}},
\end{split}
\end{equation}
where $t_2^i$ is the Armijo backtracking step size. The retraction operation is displayed in Fig. \ref{fig:RMO:c}.

With these steps, the RMO algorithm is summarized as Algorithm 1, where the convergence is proved in \cite{Absil2009}.
\begin{algorithm}[h]
\caption{The RMO Algorithm for Problem \eqref{eq:rephase}.}
\begin{algorithmic}[1]
\STATE{Initialize a point ${{\boldsymbol{\theta}}_1^0}$ and set the convergence accuracy $\epsilon_\theta$, calculate ${{\bf{d}}^0} =  - {\rm{gra}}{{\rm{d}}_{{{\boldsymbol{\theta}}_1^0}}}f$, set $i=0$;}
\STATE{{\bf{repeat}}
  \begin{enumerate}[a)]
     \item Choose the Armijo line search step $t_2^i$;
     \item Find ${{\boldsymbol{\theta}}_1^{i+1}}$ using the retract operation: ${\left[ {{{\boldsymbol{\theta}}_1^{i+1}}} \right]_m} = {{{{\left[ {{{\boldsymbol{\theta}}_1^i} + t_2^i{{\bf{d}}^i}} \right]}_m}} \mathord{\left/
 {\vphantom {{{{\left[ {{{\boldsymbol{\theta}}_1^i} + t_2^i{{\bf{d}}^i}} \right]}_m}} {\left| {{{\left[ {{{\boldsymbol{\theta}}_1^i}+ t_2^i{{\bf{d}}^i}} \right]}_m}} \right|}}} \right.
 \kern-\nulldelimiterspace} {\left| {{{\left[ {{{\boldsymbol{\theta}}_1^i} + t_2^i{{\bf{d}}^i}} \right]}_m}} \right|}}$;
     \item Determine the Riemannian gradient ${\rm{gra}}{{\rm{d}}_{{{\boldsymbol{\theta}}_1^{i+1}}}}f$ in \eqref{eq:Rieg};
     \item Calculate transport ${{\cal T}_{{{\boldsymbol{\theta}}_1^i} \mapsto {{\boldsymbol{\theta}}_1^{i+1}}}}\left( {{{\bf{d}}^i}} \right)$ according to \eqref{eq:transport};
     \item Calculate the Polak-Ribiere parameter $t_1^{i + 1}$ in \eqref{eq:Polak};
     \item Compute the conjugate direction ${{\bf{d}}^{i + 1}}$ by \eqref{eq:cgd};
  \item \(i \leftarrow i + 1\).
  \end{enumerate}}
\STATE{{\bf{until}} convergence, i.e, $\left\| {{\mathrm{gra}}{{\mathrm{d}}_{{{\boldsymbol{\theta}}_1^i}}}f} \right\| \le \epsilon_\theta$.}
\STATE{Output ${\bf{\Theta}}_1^ \star $.}
\end{algorithmic}
\end{algorithm}

Next, we focus on obtaining the optimal $\rho$. Here, we denote the corresponding optimal ${{{\boldsymbol{\theta }}_1}}$ with given $\rho$ as ${{\boldsymbol{\theta }}_1\left( \rho  \right)}$. Then, the optimal $\rho$ can be found by the following complementary slackness condition
\begin{equation}\label{eq:CSC}
\rho \sum\limits_{l = 1}^L {\left( {{\boldsymbol{\theta }}_1^H\left( \rho  \right){{\bf{U}}_l}{{\boldsymbol{\theta }}_1}\left( \rho  \right) + 2\Re \left\{ {{\boldsymbol{\theta }}_1^H\left( \rho  \right){{\bf{q}}_l}} \right\} - {{\tilde \Gamma }_l}} \right)}  = 0.
\end{equation}

For \eqref{eq:CSC}, there exists the following two cases:
\begin{itemize}
\item[1)] When $\rho=0$, if
\begin{equation}\label{eq:CSC1}
\sum\limits_{l = 1}^L {\left( {{\boldsymbol{\theta }}_1^H\left( 0  \right){{\bf{U}}_l}{{\boldsymbol{\theta }}_1}\left( 0 \right) + 2\Re \left\{ {{\boldsymbol{\theta }}_1^H\left( 0  \right){{\bf{q}}_l}} \right\} - {{\tilde \Gamma }_l}} \right)}  \le 0,
\end{equation}
then the optimal $\rho$ is $\rho=0$.

\item[2)] Otherwise, \eqref{eq:CSC} holds if and only if the following equation holds
\begin{equation}\label{eq:CSC2}
\hspace{-4mm}\gamma \left( \rho  \right) \buildrel \Delta \over = \sum\limits_{l = 1}^L {\left( {{\boldsymbol{\theta }}_1^H\left( \rho  \right){{\bf{U}}_l}{{\boldsymbol{\theta }}_1}\left( \rho  \right) + 2\Re \left\{ {{\boldsymbol{\theta }}_1^H\left( \rho  \right){{\bf{q}}_l}} \right\} - {{\tilde \Gamma }_l}} \right)}  = 0.
\end{equation}
\end{itemize}

As mentioned by \cite{PanJSAC2020} and \cite{ZhangTVT2020}, $\gamma \left( \rho  \right)$ is monotonically decreasing w.r.t. $\rho$. Thus, the bisection search method can be used to find the optimal $\rho$ for the case of $ \rho > 0$. The total procedure to find $\rho$ and ${{\boldsymbol{\theta }}_1}$ are given in Algorithm 2.
\begin{algorithm}[h]
\caption{The Algorithm for Problem \eqref{eq:rephase}.}
\begin{algorithmic}[1]
\STATE{Initialize the iterative number $t=0$, the maximum iterative number ${t_{\max }}$, error accuracy ${\epsilon_\theta }$ and ${\epsilon_\rho }$.}
\STATE{{\bf{Do while:}}
  \begin{enumerate}[a)]
     \item Given ${{\bf{W}}_k}$, calculate the objective value of \eqref{eq:rephase}, which is denote by $f\left( {{\boldsymbol{\theta }}_1^{\left( t \right)}} \right)$;
     \item Calculate $\gamma \left( 0 \right)$ by \eqref{eq:CSC1};
     \item If \eqref{eq:CSC1} holds true, the optimal value ${\rho ^{\star}} = 0$, go to Step i; Otherwise, initialize the lower and upper bounds ${\rho _l}$ and ${\rho _u}$.
     \item {\bf{while}} $\left| {{\rho _l} - {\rho _u}} \right| \ge {\epsilon _\rho }$ {\bf{do}}
     \item Calculate $\rho  = \frac{{{\rho _l} + {\rho _u}}}{2}$ and $\gamma \left( \rho  \right)$ by \eqref{eq:CSC2};
     \item If $\gamma \left( \rho  \right) \ge 0$, ${\rho _l} \leftarrow \rho $; Otherwise, ${\rho _u} \leftarrow \rho $;
      \item {\bf{end while}}
      \item Set the optimal $\rho  = \frac{{{\rho _l} + {\rho _u}}}{2}$, go to Step i;
      \item Obtain ${\boldsymbol{\theta }}_1^{\left( t \right)}$ by solving \eqref{eq:rephase} using the manifold optimization toolbox \cite{Boumal2014};
      \item Calculate the objective value $f\left( {{\boldsymbol{\theta }}_1^{\left( t+1 \right)}} \right)$ of \eqref{eq:rephase};
      \item If $t > {t_{\max }}$ or $\frac{{\left| {f\left( {{\boldsymbol{\theta }}_1^{\left( {t + 1} \right)}} \right) - f\left( {{\boldsymbol{\theta }}_1^{\left( t \right)}} \right)} \right|}}{{f\left( {{\boldsymbol{\theta }}_1^{\left( t \right)}} \right)}} \le {\epsilon_\theta }$, go to Step l; Otherwise, ${\boldsymbol{\theta }}_1^{\left( {t + 1} \right)} \leftarrow {\boldsymbol{\theta }}_1^{\left( t \right)}$, go to Step a.
      \item {\bf{return}} the optimal solution ${{\boldsymbol{\theta }}_1} = {\boldsymbol{\theta }}_1^{\left( t \right)}$.
  \end{enumerate}}
\end{algorithmic}
\end{algorithm}

It is easily known that the optimization of ${\boldsymbol{\theta }}_2$ can be obtained in a similar way, when fixing the other variables. At this time, the related parameters are given by\vspace{1.25ex}
\begin{subequations}\label{eq:relation3}
\begin{align}
&{{\bf{q}}_{1,l}} = {\rm{diag}}\left( {{{\bf{F}}_2}{{\bf{\Theta }}_1}{{\bf{F}}_1}{{\bf{W}}_l}{\bf{A}}_l^{12}{{\bf{H}}_l} + {{\bf{F}}_3}{{\bf{W}}_l}{\bf{A}}_l^{12}{{\bf{H}}_l}} \right),\vspace{1.25ex}\\
&{{\bf{q}}_{2,l}} = {\rm{diag}}\left( {{{\bf{F}}_2}{{\bf{\Theta }}_1}{{\bf{F}}_1}{\bf{\Xi F}}_1^H{\bf{\Theta }}_1^H{\bf{G}}_l^H{\bf{A}}_l^{22}{{\bf{H}}_l}} \right),\vspace{1.25ex} \\
&{{\bf{q}}_{3,l}} = {\rm{diag}}\left( {{{\bf{F}}_3}{\bf{\Xi F}}_1^H{\bf{\Theta }}_1^H{\bf{G}}_l^H{\bf{A}}_l^{22}{{\bf{H}}_l}} \right),\vspace{1.25ex}\\
\begin{split}
&{{\bf{U}}_{1,l}} = \left( {{\bf{H}}_l^H{\bf{A}}_l^{22}{{\bf{H}}_l}} \right) \odot \vspace{1.25ex}\\
& {\left( {{{\bf{F}}_2}{{\bf{\Theta }}_1}{{\bf{F}}_1}{\bf{\Xi F}}_1^H{\bf{\Theta }}_1^H{\bf{F}}_2^H + {{\bf{F}}_3}{\bf{\Xi F}}_3^H} \right)^T}, \vspace{1.25ex}\end{split}\\
&{{\bf{U}}_{2,l}} = \left( {{\bf{H}}_l^H{\bf{A}}_l^{22}{{\bf{H}}_l}} \right) \odot {\left( {{{\bf{F}}_3}{\bf{\Xi F}}_1^H{\bf{\Theta }}_1^H{\bf{F}}_2^H} \right)^T}, \vspace{1.25ex}\\
&{{\bf{U}}_{3,l}} = \left( {{\bf{H}}_l^H{\bf{A}}_l^{22}{{\bf{H}}_l}} \right) \odot {\left( {{{\bf{F}}_2}{{\bf{\Theta }}_1}{{\bf{F}}_1}{\bf{\Xi F}}_3^H} \right)^T},\vspace{1.25ex}
\end{align}
\end{subequations}
respectively.

In addition, for the discrete RC case, taking ${{\boldsymbol{\theta }}_1}$ for example, at the end of Algorithm 2, we project ${\left[ {{{\boldsymbol{\theta }}_1}} \right]_m}$ into the discrete set. In particular, we denote the solution of the two cases as ${\left[ {{{\boldsymbol{\theta }}_1}} \right]_m^{\rm{c}}}$ and ${\left[ {{{\boldsymbol{\theta }}_1}} \right]_m^{\rm{d}}}$, respectively. Then, we map ${\left[ {{{\boldsymbol{\theta }}_1}} \right]_m^{\rm{c}}}$ to obtain ${\left[ {{{\boldsymbol{\theta }}_1}} \right]_m^{\rm{d}}}$, i.e.,
${\left[ {{{\boldsymbol{\theta }}_1}} \right]_m^{\rm{d}}} = {e^{j{\phi _{{q^ \star }}}}}$, where ${q^ \star } = \mathop {\arg \min }\limits_{1 \le q \le {2^{{Q_\phi }}}} \left| {{\left[ {{{\boldsymbol{\theta }}_1}} \right]_m^{\rm{c}}} - {e^{j{\phi _q}}}} \right|$.

\subsection{Convergence and Complexity Analysis}
To this end, we finish the joint precoding and phase shifter design. The BCD algorithm is given in Algorithm 3.
\begin{algorithm}[h]
\caption{The BCD Algorithm for Problem \eqref{eq:Op}.}
\begin{algorithmic}[1]
\STATE{Initialize a feasible point $\left( {{{\bf{W}}_l^0},{\bf{\Theta }}_1^0, {\bf{\Theta }}_2^0} \right)$, and set $k=0$.}
\STATE{{\bf{repeat}}
  \begin{enumerate}[a)]
  \item Computer ${\bf{W}}_l^k$ by \eqref{eq:reRp}.
  \item Computer ${\bf{\Theta }}_1^k$ and ${\bf{\Theta }}_2^k$ by solving \eqref{eq:rephase}.
  \item Update ${{\bf{\bar H}}}_l$, ${\bf{A}}_l^{11}$, ${\bf{A}}_l^{12}$, and ${\bf{A}}_l^{22}$, respectively.
  \item \(k \leftarrow k + 1\).
  \end{enumerate}}
\STATE{{\bf{until}} some stopping criterion is satisfied. }
\STATE{Output $\left( {{{\bf{W}}_l^\star},{\bf{\Theta }}_1^\star, {\bf{\Theta }}_2^\star} \right)$.}
\end{algorithmic}
\end{algorithm}

Now, we estimate the computational complexity of Algorithm 3. Firstly, for the optimization of $\left\{ {{{\bf{W}}_l}} \right\}_{l = 1}^L$, accordingly to \cite{PanTWC2020}, the complexity of solving problem \eqref{eq:reRp} is given by ${\cal O}\left( {{{\log }_2}\left( {\frac{1}{{{\epsilon _w}}}} \right)LN_{{\rm{TX}}}^3} \right)$, where ${\epsilon _w}$ denotes the accuracy requirement.

Then, for the price mechanism-based RMO algorithm to update ${\bf{\Theta}}_1$, the complexity for calculating $\gamma \left( \rho  \right)$ is ${\cal O}\left( {M_1^2} \right)$. To obtain the optimal $\rho$, the number of iterations is ${\log _2}\left( {\frac{{{\rho _u} - {\rho _l}}}{{{\epsilon _\rho }}}} \right)$. Therefore, the total complexity is ${\cal O}\left( {{{\log }_2}\left( {\frac{{{\rho _u} - {\rho _l}}}{{{\epsilon _\rho }}}} \right)M_1^2} \right)$. Accordingly, the total complexity of Algorithm 3 is given by\vspace{1.25ex}
\begin{equation}\label{eq:comple}
\begin{split}
&C = {\cal O}\left( {\max \left\{ {{{\log }_2}\left( {\frac{1}{{{\epsilon _w}}}} \right)LN_{{\rm{TX}}}^3,} \right.} \right.\\
&\left. {\left. {{{\log }_2}\left( {\frac{{{\rho _u} - {\rho _l}}}{{{\epsilon _\rho }}}} \right)M_1^2,{{\log }_2}\left( {\frac{{{\rho _u} - {\rho _l}}}{{{\epsilon _\rho }}}} \right)M_2^2} \right\}} \right).\vspace{1.25ex}
\end{split}
\end{equation}
Hence, Algorithm 3 enjoys the polynomial time complexity, which is beneficial to implementation.

In fact, we can use the SDR method to solve problem \eqref{eq:rephase0} and then select the optimal ${\boldsymbol{\theta }}_1$ through Gaussian randomization (GR). However, according to \cite{PanTWC2020}, for the SDR-GR method, the complexity of the SDR algorithm in each iteration is given by $ {\cal O}\left( {\sqrt 2 L{{\left( {{M_1} + 1} \right)}^{0.5}}{M_1}\left( {3M_1^3 + 2M_1^2 + {M_1}} \right)} \right) \cong  {\cal O} \left( {3LM_1^{4.5} + 2LM_1^{3.5} +L M_1^{2.5}} \right)$, and the GR is with the complexity of ${\cal O}\left(L M_1^3 \right)$. Thus, the SDR-GR method leads to higher complexity than the RMO method.

In addition, for the convergence of Algorithm 3, the following Theorems hold.

\begin{theorem}
The value sequence $R\left( {{\bf{W}}_l^k,{\bf{\Theta }}_1^k,{\bf{\Theta }}_2^k} \right)$ of \eqref{eq:MM} is non-descending and can converge to a locally optimal value.
\end{theorem}
\begin{IEEEproof}
See Appendix \ref{proof1}.
\end{IEEEproof}

\begin{theorem}
The converged solution generated by Algorithm 3 is a KKT point of \eqref{eq:MM}.
\end{theorem}
\begin{IEEEproof}
See Appendix \ref{proof2}.
\end{IEEEproof}

\section{Simulation Results}\label{secSimulations}
\begin{figure}[!htb]
\captionsetup{font={small}}
\begin{center}
  \includegraphics[width=3in,angle=0]{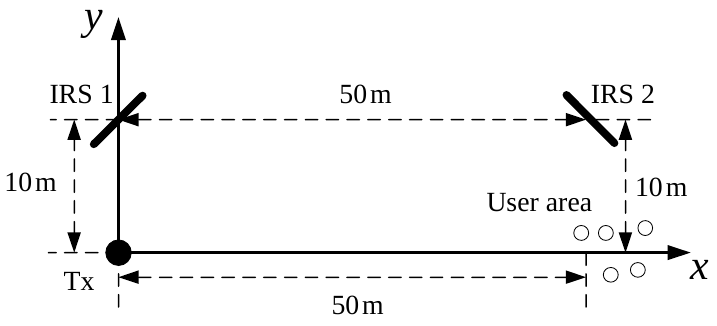}\\
  \caption{The simulation scenario.}
  \label{Fig:dep}
\end{center}
\end{figure}

Here, simulation results are provided to assess the performance of the proposed method. The scenario is shown in Fig. \ref{Fig:dep}, where there exist one Tx, two IRSs, and $L=5$ users. A three-dimensional coordinate is considered, where Tx and the IRSs are located at $\left( {0\;{\mathrm{m}},0\;{\mathrm{m}}, 10\;{\mathrm{m}}} \right)$, $\left( {0\;{\mathrm{m}},10\;{\mathrm{m}},10\;{\mathrm{m}}} \right)$, and $\left( {50\;{\mathrm{m}},10\;{\mathrm{m}},10\;{\mathrm{m}}} \right)$, respectively. Besides, the users are located in a circle centered at $\left( {50\;{\mathrm{m}},0\;{\mathrm{m}}, 2\;{\mathrm{m}}} \right)$ randomly, with radius of $5\;{\mathrm{m}}$.

Unless specified, we assume that the number of information streams of each user is $N_{d,l}= 8, \forall l$, and the weights is ${\varpi_l} = {1 \mathord{\left/
 {\vphantom {1 L}} \right.
 \kern-\nulldelimiterspace} L}, \forall l$. The numbers of antennas for Tx and users are $N_{\mathrm{TX}} = 4\times 2$ and $N_{U,l}= 2\times2,\forall l$, respectively. As for the two IRSs, we set $M_1=M_2=10\times 5$ and the number of quantized bits is ${Q_\phi }=3$. The carrier frequency is set as $28~{\rm{GHz}}$. In addition, the transmit power for Tx is set as ${P_s} = 30~{\rm{dBm}}$ and the noise power is set as ${\sigma_l ^2} = -80~{\rm{dBm}}, \forall l$ \cite{WangTWC2021}, respectively. Besides, the QoS threshold is set as $\Gamma  = 1\;{\rm{bit/s/Hz}}$ and the smooth parameter is set as $\mu=10$ \cite{Xu2001}, respectively.

In addition, we assume that the mmWave channel contains ${N_{{\rm{path}}}}=8$ propagation paths, and the distance-dependent path loss $PL\left( D \right)$ is modeled as \cite{Hong2021}:\vspace{1.25ex}
\begin{equation}\label{eq:lossmodel}
 PL\left( D \right)[{\rm{dB}}] = a  + 10b {\log _{10}}\left( D \right) + \zeta,\vspace{1.25ex}
 \end{equation}
where $D$ denotes the link distance and $\zeta  \sim {\cal N}\left( {0,{\varrho^2}} \right)$. According to the real-world channel measurements for $28\;{\rm{GHz}}$ channels \cite{AkdenizJSAC2014}, the parameters in \eqref{eq:lossmodel} are set as $a = 61.4$, $b = 2$ and $\varrho= 5.8~{\rm{dB}}$ for a LoS path, and $a  = 72.0$, $b= 2.92$ and $\varrho= 8.7~{\rm{dB}}$ for the NLoS paths, respectively. In addition, the element spacing are set to be $d = {\lambda  \mathord{\left/
 {\vphantom {\lambda  2}} \right.
 \kern-\nulldelimiterspace} 2}$.

\subsection{Convergence}
\begin{figure}[!htb]
\captionsetup{font={small}}
\begin{center}
  \includegraphics[width=3in,angle=0]{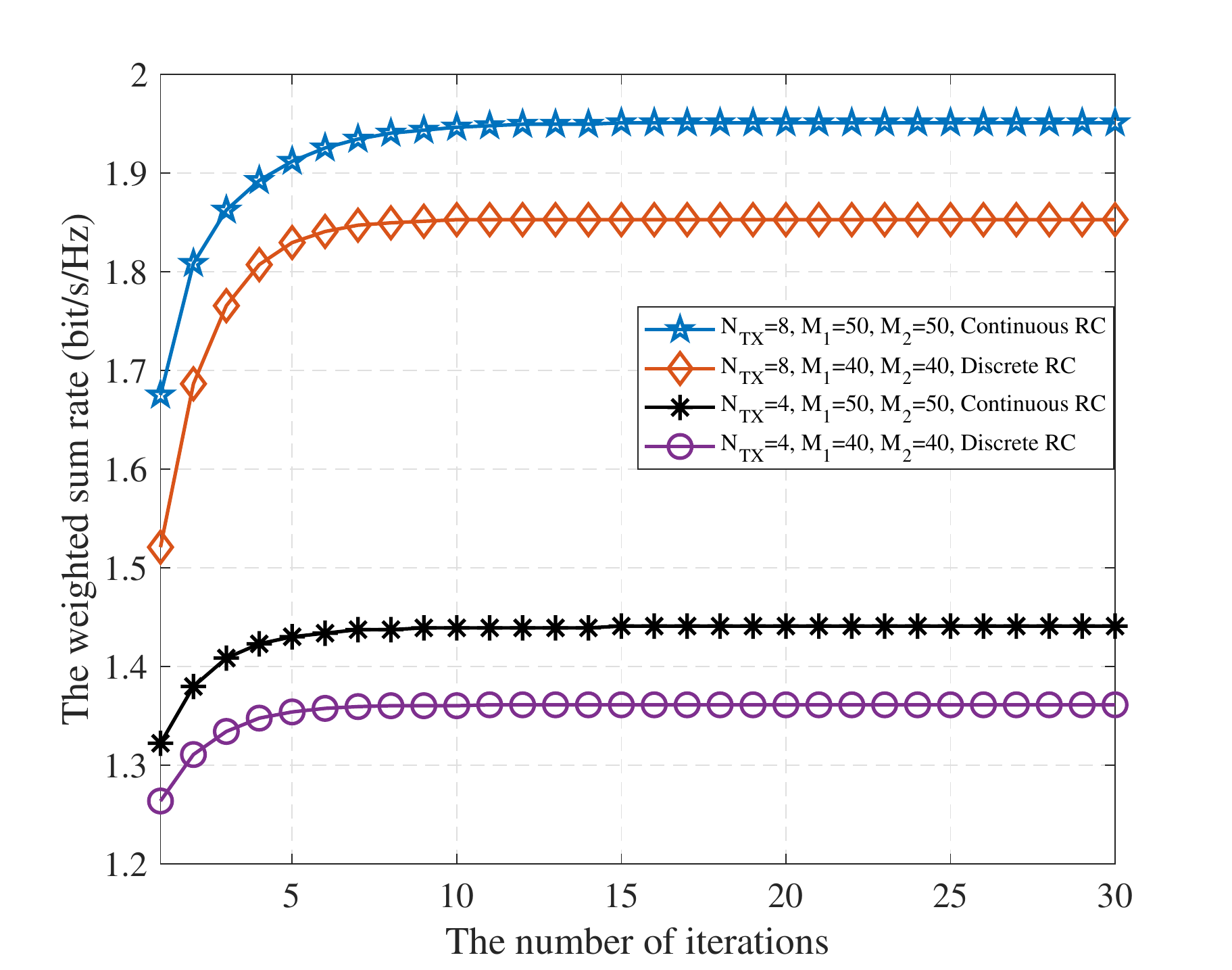}\\
  \caption{The convergence behaviour of the RMO algorithm.}
  \label{Fig:inner}
\end{center}
\end{figure}

Firstly, the convergence behaviour of the inner RMO algorithm are tested. From Fig. \ref{Fig:inner}, we can see that for different $N_{\mathrm{TX}}$, $M_1$ and $M_2$, the WSR increases with the iteration numbers, and gradually converges almost within $30$ iterations, either in continuous or discrete RC case, which demonstrates the practicality of the RMO method. Moveover, larger $N_{\mathrm{TX}}$, $M_1$ or $M_2$ leads to slower converge speed, since more variables should be optimized.

\begin{figure}[!htb]
\captionsetup{font={small}}
\begin{center}
  \includegraphics[width=3in,angle=0]{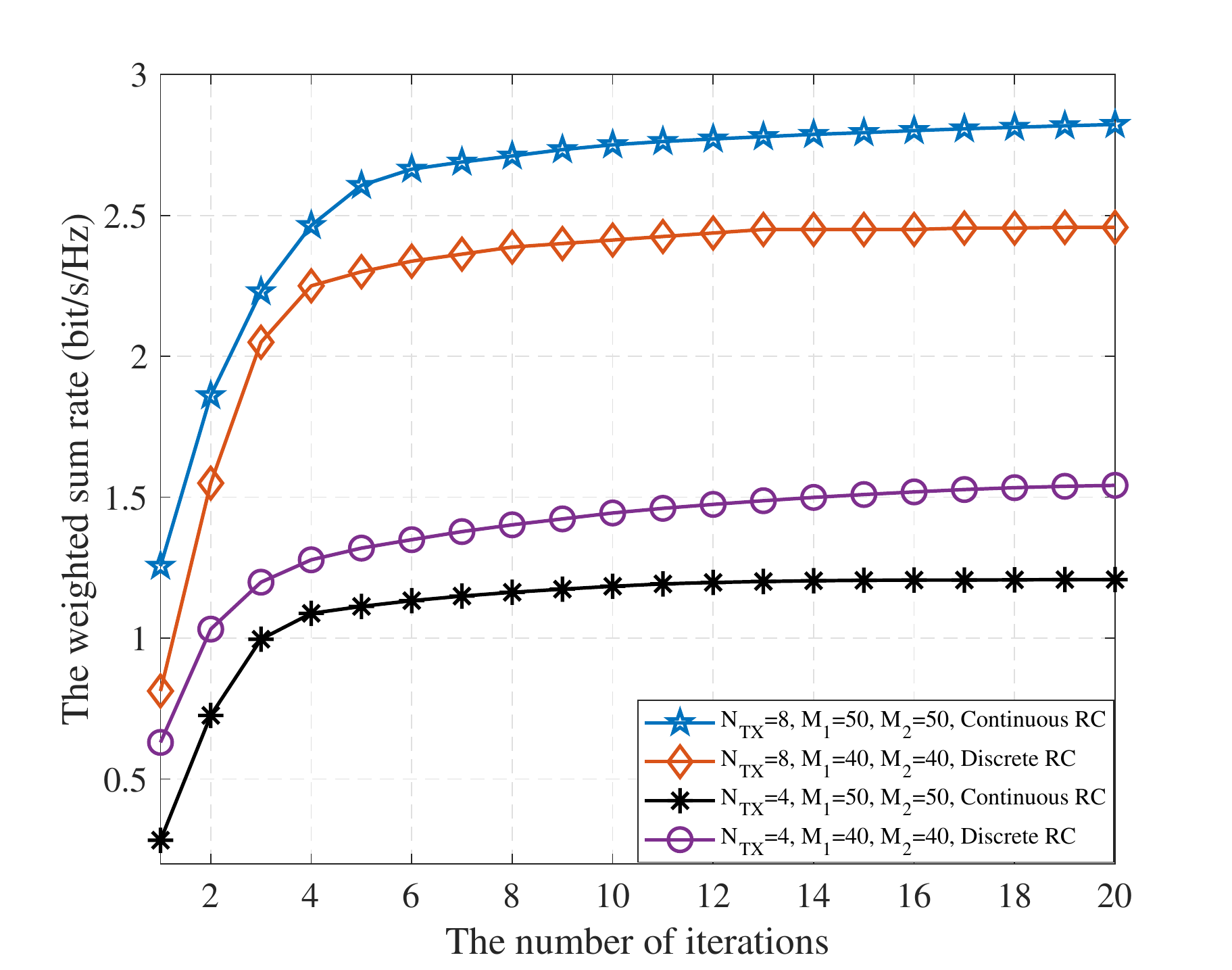}\\
  \caption{The convergence behaviour of the BCD algorithm.}
  \label{Fig:outer}
\end{center}
\end{figure}

Then, the convergence behaviour of the outer BCD algorithm are evaluated in Fig. \ref{Fig:outer}. From Fig. \ref{Fig:outer}, we can find that the WSR increases with the number of iterations, and gradually converges almost in $20$ iterations for various $N_{\mathrm{TX}}$, $M_1$ and $M_2$, which confirms the efficiency of the BCD method.

\subsection{Performance Evaluation}
Now, we study the impact of the main system parameters on the WSR. To show the effectiveness of the adopted double-IRS design, we compare the proposed method with several benchmarks: 1) random double-IRS design, e.g., select the phase shift randomly; 2) a single-IRS near the Tx; 3) a single-IRS near the users; and 4) The SDR-GR method to design the phase shifters. These approaches are labelled as ``Continuous RC", ``Discrete RC", ``Random IRS", ``Near Tx", ``Near the users", and ``SDR-GR method", respectively.

\begin{figure}[!htb]
\captionsetup{font={small}}
\begin{center}
  \includegraphics[width=3in,angle=0]{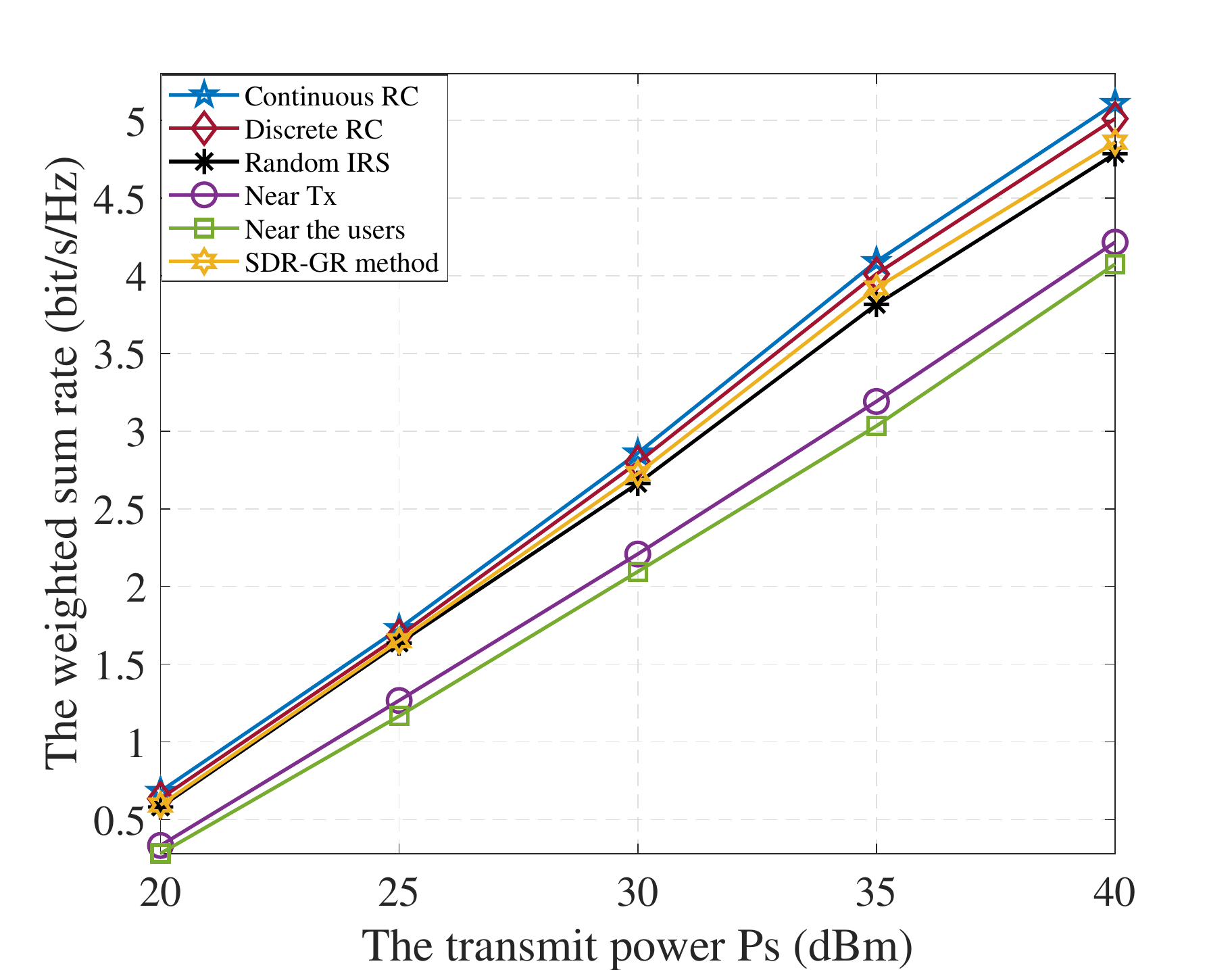}\\
  \caption{The WSR versus the transmit power $P_s$.}
  \label{Fig:Ps}
\end{center}
\end{figure}

Firstly, we show the WSR versus the transmit power \(P_s\) in Fig. \ref{Fig:Ps}, where the number of elements for single-IRS-enabled schemes is $M = {M_1} + {M_2} = 100$. From this figure, we can see that whether in continuous RC case or discrete RC case, the double-IRS design outperforms the other designs, while the single-IRS near the users case suffers the worst performance. This is mainly due to the fact that the reflected signals through the double-reflecting link and single-reflecting link can be constructively added at users by the proposed method, thus is beneficial to improve the WSR. In addition, the performance gap between the continuous RC case or discrete RC case is relatively small, which suggests the practicability of the proposed algorithm. Moreover, we can observe from Fig. \ref{Fig:Ps} that the proposed method obtains better performance than the SDR-GR method, due to the proposed optimization framework.

\begin{figure}[!htb]
\captionsetup{font={small}}
\begin{center}
  \includegraphics[width=3in,angle=0]{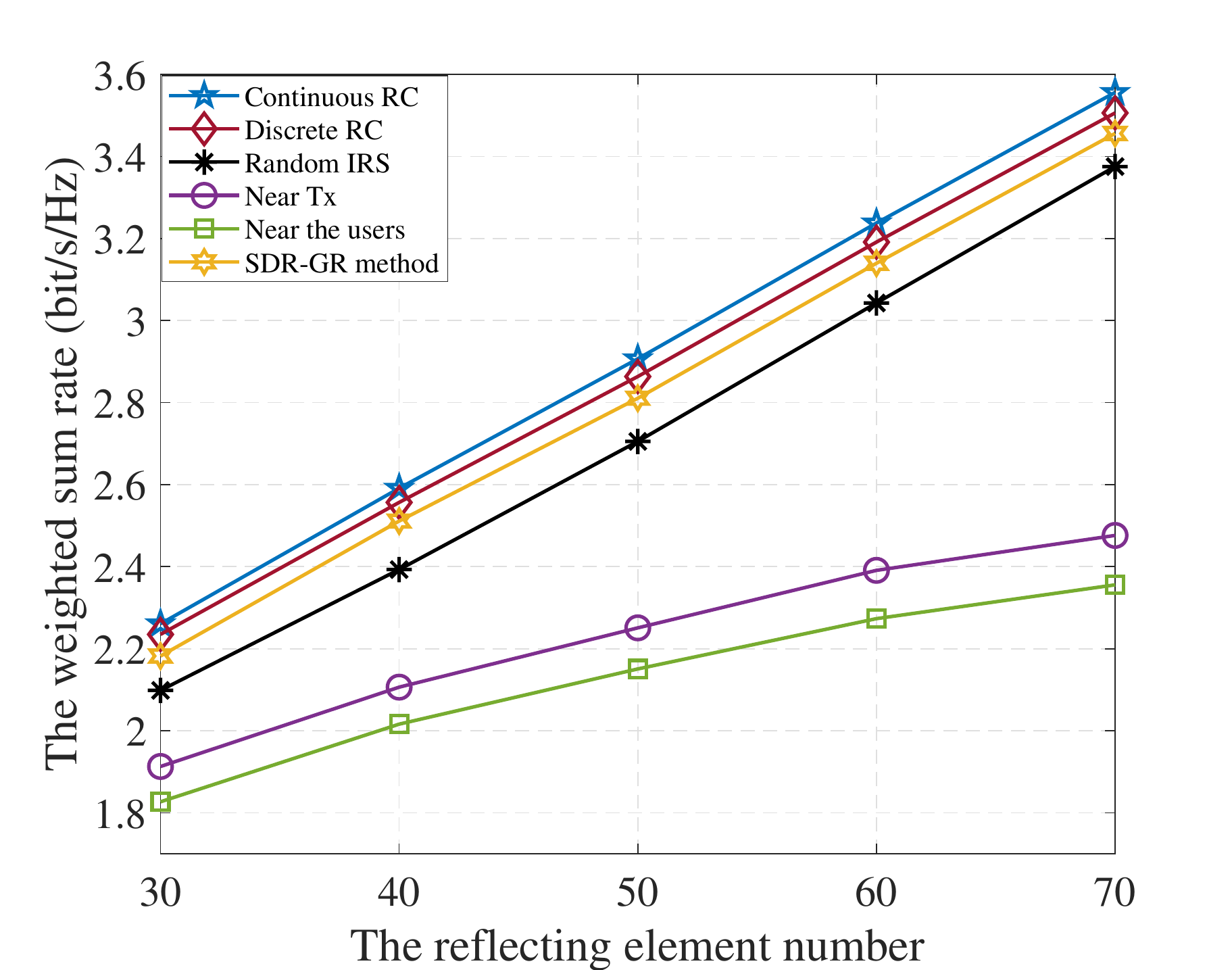}\\
  \caption{The WSR versus the number of reflection elements $M$.}
  \label{Fig:M}
\end{center}
\end{figure}

Then, we show the WSR versus the number of the reflection elements in Fig. \ref{Fig:M}. The $x$-axis in Fig. \ref{Fig:M} denotes the total number of elements for single-IRS case, or the number of elements for each IRS in double-IRS case. From Fig. \ref{Fig:M}, we can see that for all these methods, the WSR increases with \(M\). The result comes from two folds. Firstly, with a larger \(M\), the incident power at the IRS is enhanced, thus a higher array gain can be obtained. Secondly, with larger \(M\), the reflected signal power received by the users increases, when the RC are optimized properly. This result suggests that a larger IRS can improve the rate performance with proper phase shift. In addition, from Fig. \ref{Fig:M}, we can see that the scaling order of double-IRS case is higher than the scaling order of single-IRS, which indicates the superiority of double-IRS.

\begin{figure}[!htb]
\captionsetup{font={small}}
\begin{center}
  \includegraphics[width=3in,angle=0]{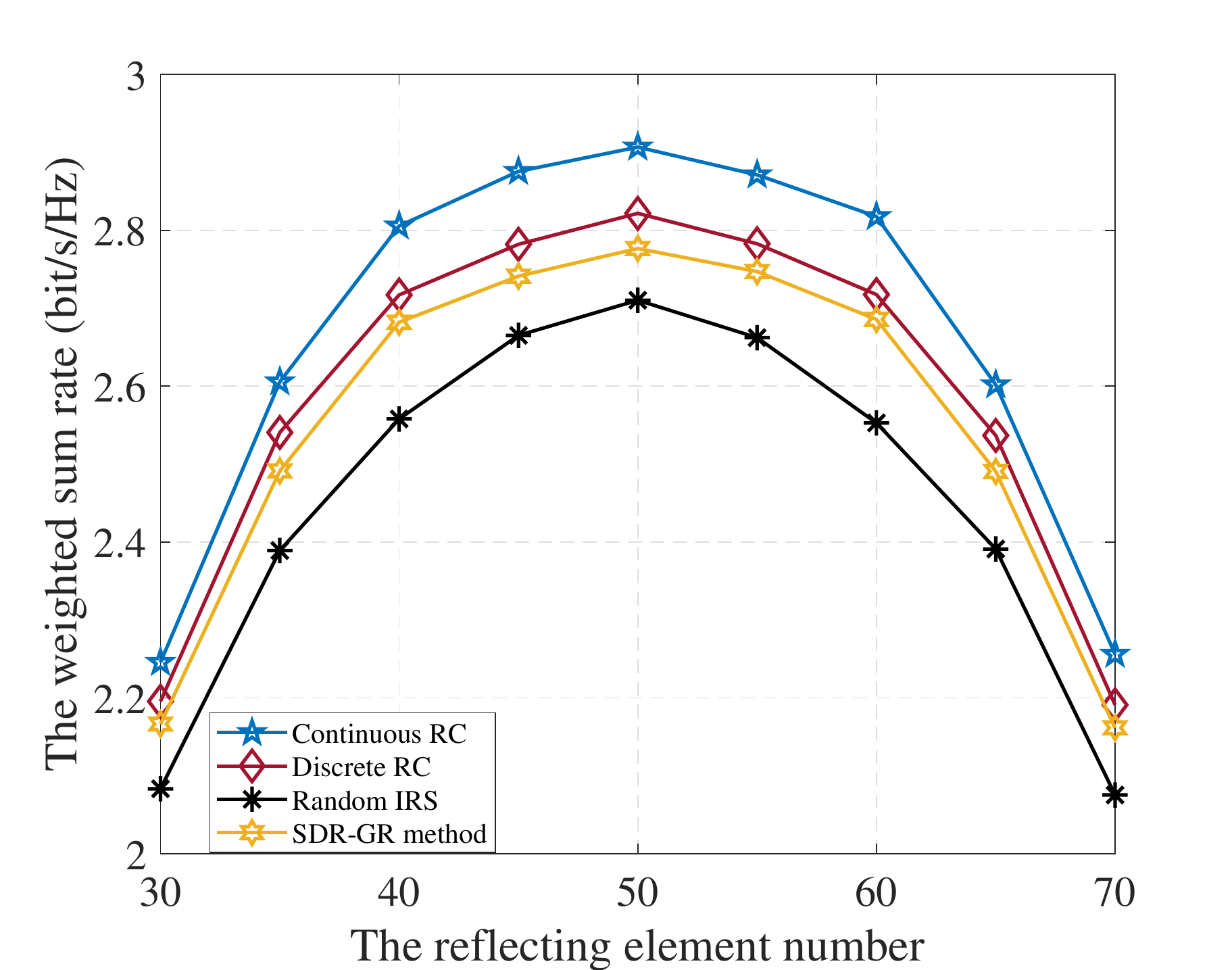}\\
  \caption{The WSR versus the number of reflection elements $M_1$.}
  \label{Fig:M1}
\end{center}
\end{figure}

Next, in Fig. \ref{Fig:M1}, we show the WSR versus the number of the reflection elements $M_1$, with fixed total number of elements $M = {M_1} + {M_2} = 100$. It can be observed in Fig. \ref{Fig:M1} that the WSR obtains the maximized value when the two IRSs have nearly equal number of elements. This is mainly due to that equally assigned the IRS elements can effectively balance the passive BF gains between the two single-reflection links, thus is beneficial to improve the WSR.

\begin{figure}[!htb]
\captionsetup{font={small}}
\begin{center}
  \includegraphics[width=3in,angle=0]{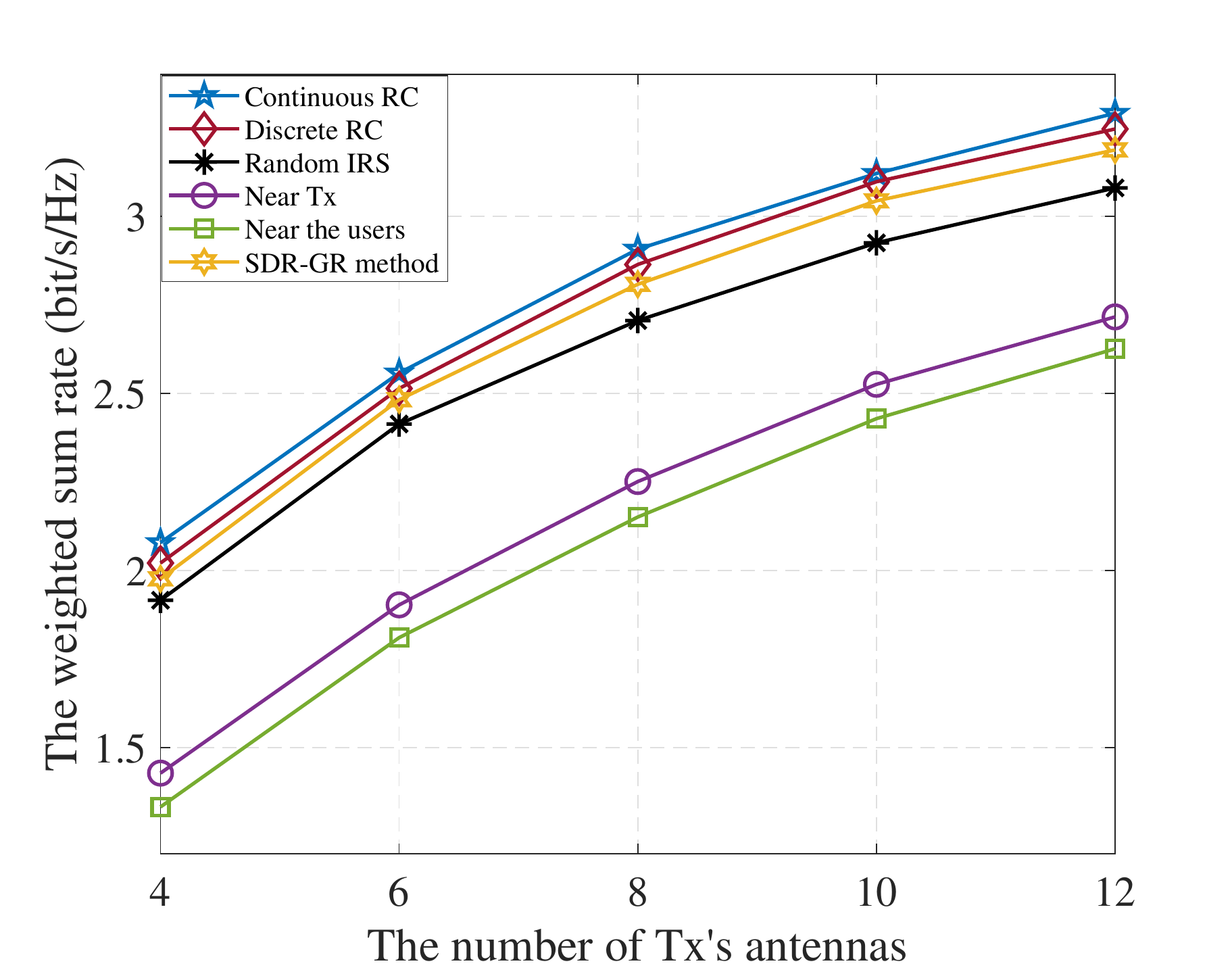}\\
  \caption{The WSR versus the number of Tx's antennas.}
  \label{Fig:Ntx}
\end{center}
\end{figure}

\begin{figure}[!htb]
\captionsetup{font={small}}
\begin{center}
  \includegraphics[width=3in,angle=0]{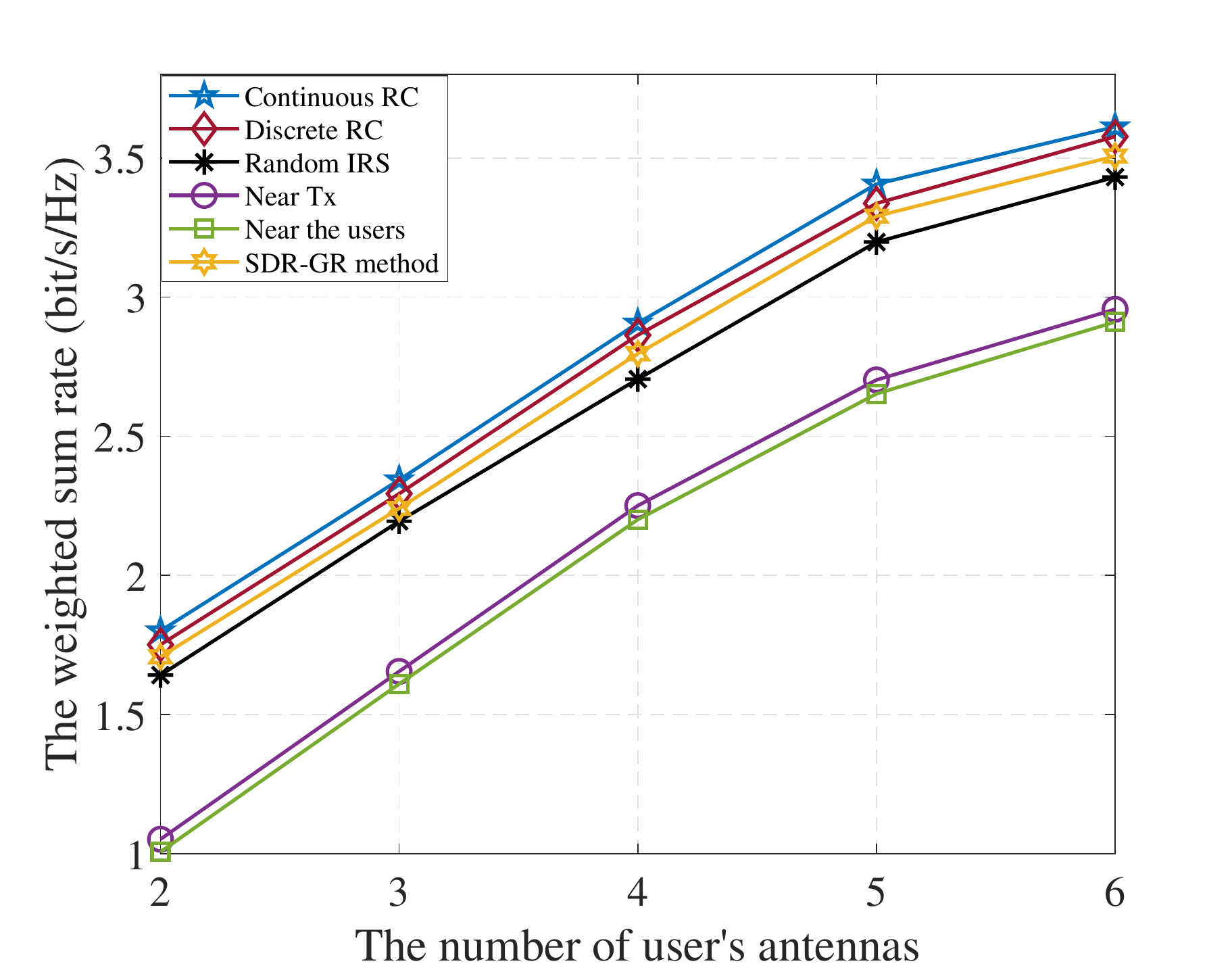}\\
  \caption{The WSR versus the number of user's antennas.}
  \label{Fig:Nul}
\end{center}
\end{figure}

Moreover, we show the WSR versus the number of Tx's antennas $N_{\mathrm{TX}} $ and the number of user's antennas $N_{U,l}$ in Fig. \ref{Fig:Ntx} and Fig. \ref{Fig:Nul}, respectively. From these two figures, we can see that for all these methods, the WSR increases with $N_{\mathrm{TX}}$ or $N_{U,l}$, since more spatial degrees of freedom are available for effective precoding and decoding with more antennas.

\begin{figure}[!htb]
\captionsetup{font={small}}
\begin{center}
  \includegraphics[width=3in,angle=0]{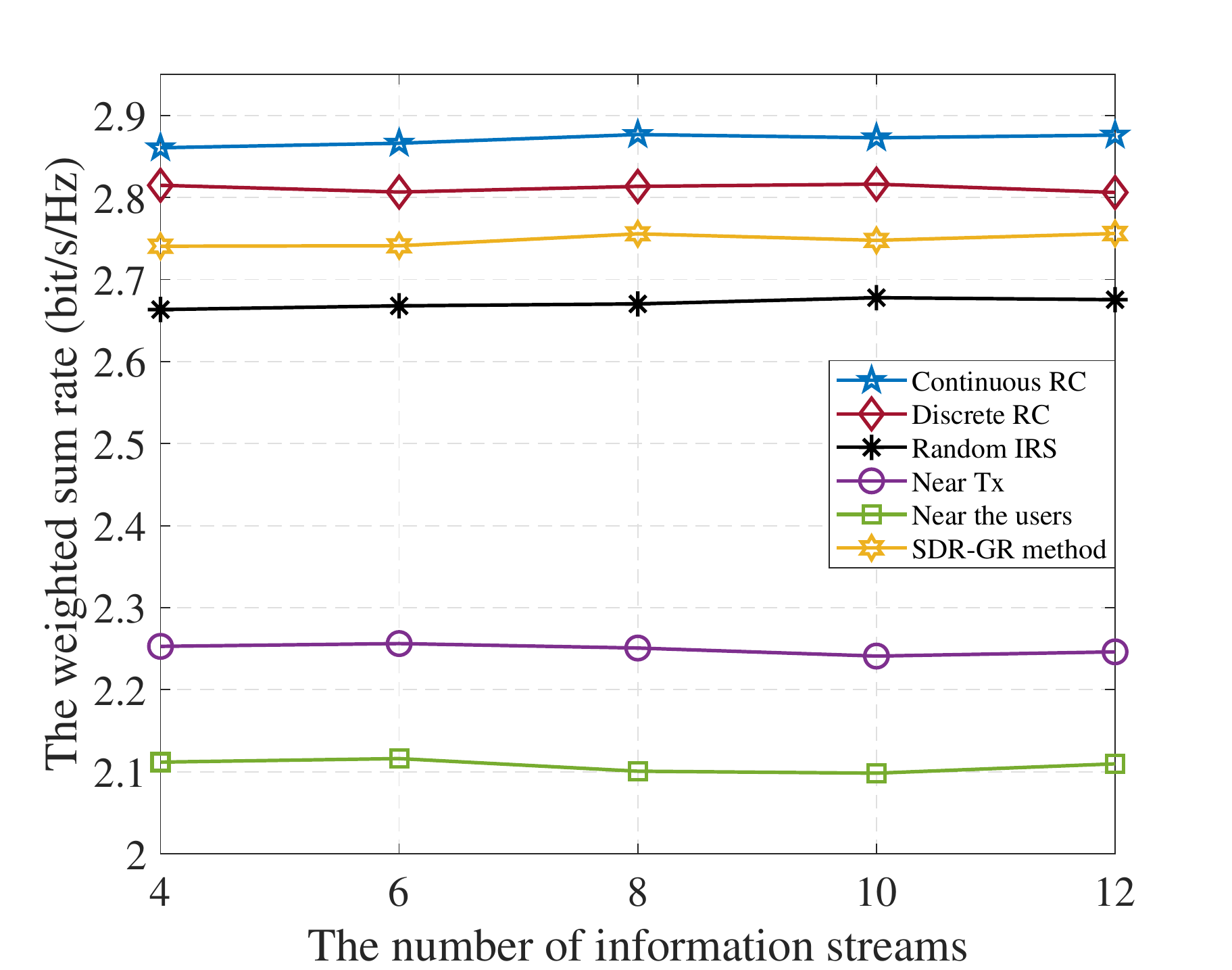}\\
  \caption{The WSR versus the number of information streams.}
  \label{Fig:Ndl}
\end{center}
\end{figure}

Then, we show the WSR versus the number of information streams $N_{d,l}$ in Fig. \ref{Fig:Ndl}. Here, since the total transmit power is a fixed value, thus the power of each symbol decreases with the increase of $N_{d,l}$. Therefore, the obtained WSRs of these methods are almost constants with different $N_{d,l}$.

\begin{figure}[!htb]
\captionsetup{font={small}}
\begin{center}
  \includegraphics[width=3in,angle=0]{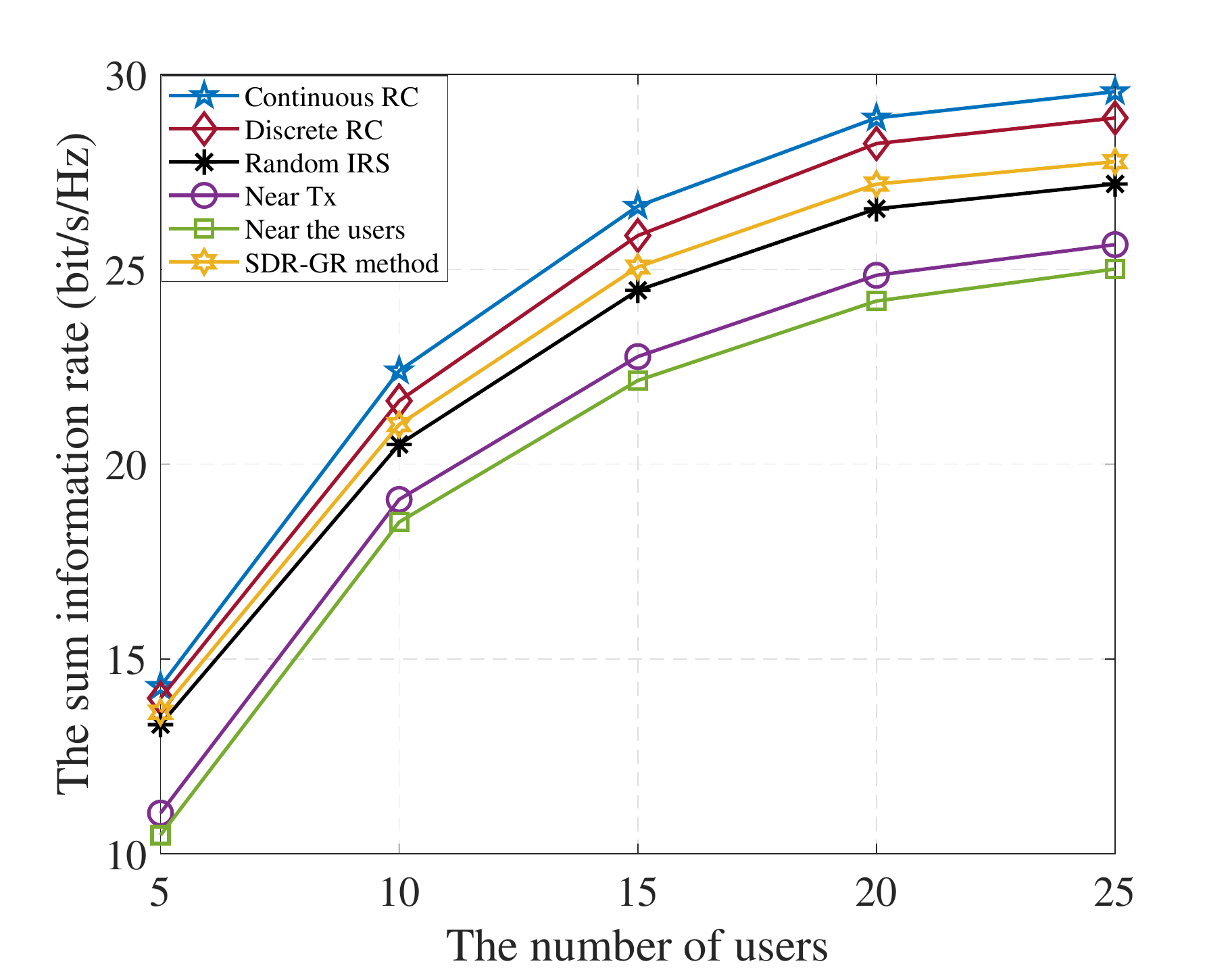}\\
  \caption{The WSR versus the number of users.}
  \label{Fig:usernum}
\end{center}
\end{figure}

Lastly, in Fig. \ref{Fig:usernum}, we show the sum information rate versus the number of users $L$. From Fig. \ref{Fig:usernum}, one can see that the rate increases with $L$, but with a marginal return. This is mainly due to the fact that with larger $L$, the mutual interference between users becomes the main bottleneck of the performance, hence limiting the rate growth. In this case, some other techniques are needed to further improve the spectral efficiency.

\section{Conclusion}\label{secConclusions}
This paper proposed a new hybrid precoding and phase shift design in mmWave MIMO network aided by two cascaded IRSs. Due to the non-convexity of the WSR maximization objective and the minimum information rate constraint, we first reformulated it into a tractable one by the MM method, then, a BCD algorithm was proposed, where the variables were optimized in an alternating manner. Specifically, the digital precoding was solved by a QCQP, while the phase shifters were solved by the price-based RMO method. Simulation results verified the advantages of the proposed method and the efficiency of double-IRS in improving the spectral efficiency.

\appendices
\section{Proof of Theorem 1}\label{proof1}
Here, we denote the objective function of \eqref{eq:MM} as $R\left( {{{\bf{W}}_l},{{\bf{\Theta }}_1},{{\bf{\Theta }}_2}} \right)$. Also, we denote $\left( {{\bf{W}}_l^k,{\bf{\Theta }}_1^k,{\bf{\Theta }}_2^k} \right)$ as the obtained solution at the $k$-th iteration in Step 2 of Algorithm 3. Then, the following relation holds:\vspace{1.25ex}
\begin{equation}\label{eq:proof1}
\begin{split}
& R\left( {{\bf{W}}_l^k,{\bf{\Theta }}_1^k,{\bf{\Theta }}_2^k,{\bf{A}}_l^{11,k},{\bf{A}}_l^{12,k},{\bf{A}}_l^{22,k}} \right)\\
&\le R\left( {{\bf{W}}_l^{k + 1},{\bf{\Theta }}_1^k,{\bf{\Theta }}_2^k,{\bf{A}}_l^{11,k},{\bf{A}}_l^{12,k},{\bf{A}}_l^{22,k}} \right)\\
& \le R\left( {{\bf{W}}_l^{k + 1},{\bf{\Theta }}_1^{k + 1},{\bf{\Theta }}_2^k,{\bf{A}}_l^{11,k},{\bf{A}}_l^{12,k},{\bf{A}}_l^{22,k}} \right)\\
& \le R\left( {{\bf{W}}_l^{k + 1},{\bf{\Theta }}_1^{k + 1},{\bf{\Theta }}_2^{k + 1},{\bf{A}}_l^{11,k},{\bf{A}}_l^{12,k},{\bf{A}}_l^{22,k}} \right)\\
& \le R\left( {{\bf{W}}_l^{k + 1},{\bf{\Theta }}_1^{k + 1},{\bf{\Theta }}_2^{k + 1},{\bf{A}}_l^{11,k + 1},{\bf{A}}_l^{12,k + 1},{\bf{A}}_l^{22,k + 1}} \right),
\end{split}
\end{equation}
where the first to the third inequality is due to Step 2-a and Step 2-b in Algorithm 3, and the last inequality is due to Step 2-c in Algorithm 3, respectively. Thus, \eqref{eq:proof1} generates a monotonically increasing sequence $R\left( {{\bf{W}}_l^k,{\bf{\Theta }}_1^k,{\bf{\Theta }}_2^k} \right)$. Moreover, $R\left( {{\bf{W}}_l^k,{\bf{\Theta }}_1^k,{\bf{\Theta }}_2^k} \right)$ is upper-bounded due to \eqref{eq:Opc1}, thus guarantee to converge.

\section{Proof of Theorem 2}\label{proof2}
As mentioned, the sequence $R\left( {{\bf{W}}_l^k,{\bf{\Theta }}_1^k,{\bf{\Theta }}_2^k} \right)$ will converges to $R\left( {{\bf{W}}_l^ \star ,{\bf{\Theta }}_1^ \star ,{\bf{\Theta }}_2^ \star } \right)$ as $k \to \infty $.

Then, when $k \to \infty $, the KKT conditions are given as:
\begin{subequations}\label{eq:LagKKT1}
\begin{align}
&\left( {u_0^ \star {\bf{I}} + \left( {u_l^ \star  + {\varpi _l}} \right){{\bf{V}}_l}} \right){\bf{w}}_l^ \star  + \left( {u_l^ \star  + {\varpi _l}} \right){{\bf{v}}_l} = {\bf{0}},\forall l, \label{eq:LagKKT11}\\
&~{u_0^ \star }\left( {\sum\limits_{l = 1}^L {{\bf{w}}_l^ \star {{\left( {{\bf{w}}_l^ \star } \right)}^H} - {P_s}} } \right) = 0, \label{eq:LagKKT12}\\
&~u_l^ \star \left( {{\bf{w}}_l^H{{\bf{V}}_l}{{\bf{w}}_l} + 2\Re \left\{ {{\bf{v}}_l^H{{\bf{w}}_l}} \right\} + {{\tilde \Gamma }_l}} \right) = 0, \label{eq:LagKKT13}
\end{align}
\end{subequations}
where ${{u_0} \ge 0}$ is the dual variable for \eqref{eq:reRpc1} and $\left\{ {{u_l} \ge 0} \right\}_{l = 1}^L$ are the dual variables for \eqref{eq:reRpc2}, respectively.

Besides, the Lagrange function of \eqref{eq:rephase} is
\begin{equation}\label{eq:LagKKTtheta}
\begin{split}
&{\cal L}\left( {{{\boldsymbol{\theta }}_1},{\boldsymbol{\eta }}} \right) =\sum\limits_{l = 1}^L {{\varpi _l}\left( {{\boldsymbol{\theta }}_1^H{{\bf{U}}_l}{{\boldsymbol{\theta }}_1} + 2\Re \left\{ {{\boldsymbol{\theta }}_1^H{{\bf{q}}_l}} \right\}} \right)} \\
& ~~~~~~~~~~+ \rho \left( {\sum\limits_{l = 1}^L {{e^{\mu \left( {{\boldsymbol{\theta }}_1^H{{\bf{U}}_l}{{\boldsymbol{\theta }}_1} + 2\Re \left\{ {{\boldsymbol{\theta }}_1^H{{\bf{q}}_l}} \right\}} \right)}}}  - {e^{\mu {{\tilde \Gamma }_l}}}} \right)\\
& ~~~~~~~~~~+ \sum\limits_{m = 1}^{{M_1}} {{\eta _m}} \left( {\left[ {{{\boldsymbol{\theta }}_1}} \right]_m^ * {{\left[ {{{\boldsymbol{\theta }}_1}} \right]}_m} - 1} \right),
\end{split}
\end{equation}
where ${\boldsymbol{\eta }} \ge {\bf{0}} \in {\mathbb{R}^{{M_1} \times 1}}$ is the dual variable w.r.t. the UMC.

Then, the KKT condition is\vspace{1.25ex}
\begin{subequations}\label{eq:LagKKT2}
\begin{align}
&{\left[ {{\bf{U}}{\boldsymbol{\theta }}_1^ \star  + {\bf{q}}} \right]_m} + \eta _m^ \star {\left[ {{\boldsymbol{\theta }}_1^ \star } \right]_m} = 0,\forall m, \vspace{1.25ex}\label{eq:LagKKT21} \\
&\eta _m^ \star \left( {\left[ {{\boldsymbol{\theta }}_1^ \star } \right]_m {{\left[ {{\boldsymbol{\theta }}_1^ \star } \right]}_m ^ *} - 1} \right) = 0,\forall m. \label{eq:LagKKT22}\vspace{1.25ex}
\end{align}
\end{subequations}

Similarly proof can be obtained for ${{{\boldsymbol{\theta }}_2}}$. Since $\left\{ {{\bf{w}}_l^ \star } \right\}_{l = 1}^L$, ${{\boldsymbol{\theta }}_1^ \star }$ and ${{\boldsymbol{\theta }}_2^ \star }$ are the optimal solutions to \eqref{eq:reRp} and \eqref{eq:rephase}, respectively, then the KKT conditions holds. Thus, $\left( {\left\{ {{\bf{w}}_l^ \star } \right\}_{l = 1}^L,{\boldsymbol{\theta }}_1^ \star ,{\boldsymbol{\theta }}_2^ \star } \right)$ is a KKT point.

\section{Introduction of MM and MO methods}
\begin{figure}[!htb]
\captionsetup{font={small}}
\begin{center}
  \includegraphics[width=2.4in,angle=0]{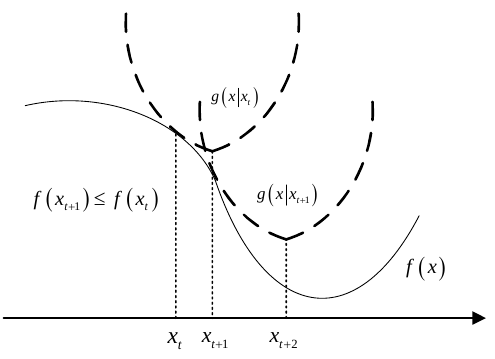}\\
  \caption{The MM procedure.}
  \label{Fig:mm}
\end{center}
\end{figure}

The MM procedure consists of two steps. In the first majorization step, we aim to develop a surrogate function that locally approximates the objective function with their difference minimized at the current point. Then, in the minimization step, we minimize the surrogate function \cite{SunTSP2017}. To be specific, let us consider the following optimization problem \vspace{1.25ex}
\begin{subequations}\label{eq:MMmethod}
\begin{align}
&\mathop {\min }\limits_{\bf{x}} \;f\left( {\bf{x}} \right) \label{eq:MMmethodo} \\
&\;\;{\rm{s.t.}}\;\;{\bf{x}} \in {\cal X}, \label{eq:MMmethodc1} \vspace{1.25ex}
\end{align}
\end{subequations}
where ${\cal X}$ is a nonempty closed set and $f\left(\cdot  \right)$ is a continuous function. Initialized as ${{\bf{x}}_0} \in {\cal X}$, MM generates a sequence of feasible points ${{\bf{x}}_t}$ by the following induction. At point ${{{\bf{x}}_t}}$, in the majorization step we construct a continuous surrogate function $g\left( {\cdot\left| {{{\bf{x}}_t}} \right.} \right)$ satisfying the upper bound property that $g\left( {{\bf{x}}\left| {{{\bf{x}}_t}} \right.} \right) \ge f\left( {\bf{x}} \right) + {c_t},\forall {\bf{x}} \in {\cal X}$, where ${c_t} = g\left( {{{\bf{x}}_t}\left| {{{\bf{x}}_t}} \right.} \right) - f\left( {{{\bf{x}}_t}} \right)$. Thus, the difference of $g\left( {\cdot\left| {{{\bf{x}}_t}} \right.} \right)$ and $f\left(\cdot  \right)$ is minimized at ${{{\bf{x}}_t}}$. Then in the minimization step, we update ${\bf{x}}$ as ${{\bf{x}}_{t + 1}} \in \mathop {\arg \min }\limits_{{\bf{x}} \in {\cal X}} g\left( {{\bf{x}}\left| {{{\bf{x}}_t}} \right.} \right)$. The procedure is shown in Fig. \ref{Fig:mm}.

In the MO approach, as shown in Fig. \ref{fig:RMO}, the main procedure consists of three steps: 1) calculating the gradient; 2) transporting the vector; and 3) retraction. In fact, the MO method is similar to the gradient-based optimization in the Euclidean space. However, we need to determine which kind of manifolds can be utilized, based on specific constraints. For the UMC associated with the phase shifts, the commonly used manifolds are the Riemannian manifold or the complex circle manifold \cite{AlhujailiTSP2019}.

\ifCLASSOPTIONcaptionsoff
  \newpage
\fi

\bibliographystyle{IEEEtran}
\bibliography{IEEEabrv,mybib}

\end{document}